%% file: anonymous-submission-latex-2024.tex
\documentclass[letterpaper]{article} % DO NOT CHANGE THIS
\usepackage{aaai24}  % DO NOT CHANGE THIS
\usepackage{times}  % DO NOT CHANGE THIS
\usepackage{helvet}  % DO NOT CHANGE THIS
\usepackage{courier}  % DO NOT CHANGE THIS
\usepackage[hyphens]{url}  % DO NOT CHANGE THIS
\usepackage{graphicx} % DO NOT CHANGE THIS
\usepackage{natbib}  % DO NOT CHANGE THIS AND DO NOT ADD ANY OPTIONS TO IT
\usepackage{caption} % DO NOT CHANGE THIS AND DO NOT ADD ANY OPTIONS TO IT
\usepackage{algorithm}
\usepackage{algorithmic}

\usepackage{newfloat}
\usepackage{listings}
\DeclareCaptionStyle{ruled}{labelfont=normalfont,labelsep=colon,strut=off} % DO NOT CHANGE THIS
\floatstyle{ruled}
\newfloat{listing}{tb}{lst}{}
\floatname{listing}{Listing}
\usepackage[toc,page]{appendix}

\newcommand{\squishlist}{
   \begin{list}{\small $\bullet$}
    { \setlength{\itemsep}{0pt}      \setlength{\parsep}{1pt}
      \setlength{\topsep}{1pt}       \setlength{\partopsep}{1pt}
     \setlength{\leftmargin}{1.2em} \setlength{\labelwidth}{1em}
      \setlength{\labelsep}{0.5em} } }
\newcommand{\squishend}{  \end{list}  }
\newcommand{\ignore}[1]{}

\newcommand{\tabincell}[2]{\begin{tabular}{@{}#1@{}}#2\end{tabular}}

\usepackage{enumitem}
\usepackage{xcolor}
\usepackage{afterpage}
\usepackage{placeins}
\usepackage{graphicx}

\usepackage{tabularx}
\usepackage{multirow}
\usepackage{caption}
\usepackage{booktabs}
\usepackage{subcaption}
\usepackage{makecell}
\usepackage{adjustbox}
\usepackage{array}    % For table wrapping
\usepackage[flushleft]{threeparttable} % For adding notes below the table
\usepackage{placeins}
\usepackage{caption}
\usepackage{alphalph}

\usepackage{amsmath}
\usepackage{tikz}
\usetikzlibrary{automata, positioning, arrows}
\newcolumntype{P}[1]{>{\centering\arraybackslash}p{#1}}
\usepackage{bibentry}

\title{Understanding Decision Subjects' Engagement with and Perceived Fairness of AI Models When Opportunities of Qualification Improvement Exist}

\author{
    Meric Altug Gemalmaz\textsuperscript{\rm 1},
    Ming Yin\textsuperscript{\rm 1}
}
\affiliations {
    \textsuperscript{\rm 1}Purdue University, West Lafayette, Indiana, USA\\
    mgemalma@purdue.edu, mingyin@purdue.edu
}

\nocopyright

\begin{document}

\maketitle

%\title{Exploring Fairness Effects on Decision Subjects' Improvement in Repeated Interactions with AI-Based Decision Systems}


% REMOVE THIS: bibentry
% This is only needed to show inline citations in the guidelines document. You should not need it and can safely delete it.

% END REMOVE bibentry

\begin{abstract}
We explore how an AI model's decision fairness affects people's engagement with and perceived fairness of the model if they are subject to its decisions, but could repeatedly and strategically respond to these decisions. Two types of strategic responses are considered---people could determine whether to continue interacting with the model, and whether to invest in themselves to improve their chance of future favorable decisions from the model. Via three human-subject experiments, we found that in decision subjects' strategic, repeated interactions with an AI model, the model's decision fairness does not change their willingness to interact with the model or to improve themselves, even when the model exhibits unfairness on salient protected attributes. However, decision subjects still perceive the AI model to be less fair when it systematically biases against their group, especially if the difficulty of improving one's qualification for the favorable decision is larger for the lowly-qualified people.
\end{abstract}

% <concept_id>10003120.10003121.10011748</concept_id>
% <concept_desc>Human-centered computing~Empirical studies in HCI</concept_desc>
% <concept_id>10010147.10010257</concept_id>
% <concept_desc>Computing methodologies~Machine learning</concept_desc>

% \ccsdesc[500]{Human-centered computing~Empirical studies in HCI}
% \ccsdesc[500]{Computing methodologies~Machine learning}
% %%
% %% Keywords. The author(s) should pick words that accurately describe
% %% the work being presented. Separate the keywords with commas.
% \keywords{AI-based decision systems, fairness, perceived fairness, retention, qualification improvement, human-subject experiments, human-AI interaction}

\section{Introduction}
\input{intro}

\section{Related Work}

\input{related}

\section{Study 1}

\input{study1}

\section{Study 2%: When the Qualification Improvement Difficulty Varies with Current Qualification Levels
}

\input{study2}

\section{Study 3%: When AI Fairness is Examined On Protected Attributes
}

\input{study3}

\section{Conclusions and Discussions}
\input{discussion}

\bibliography{anonymous-submission-latex-2024}

\clearpage
\newpage

\begin{appendices}
%\appendix
\setcounter{secnumdepth}{1} 
\newcommand{\hbAppendixPrefix}{A}
\renewcommand{\thefigure}{\hbAppendixPrefix\arabic{figure}}
\setcounter{figure}{0}
\renewcommand{\thetable}{\hbAppendixPrefix\arabic{table}} 
\setcounter{table}{0}
\input{Appendix}

\end{appendices}

% \begin{appendices}
% % Define a new counter for appendices
% \newcounter{AppendixCounter}
% \renewcommand{\theAppendixCounter}{\Alph{AppendixCounter}}

% % Redefine the section command to include the custom appendix counter
% \let\oldsection\section
% \renewcommand{\section}[1]{%
%     \stepcounter{AppendixCounter}%
%     \oldsection{Appendix \theAppendixCounter\hspace{1em}#1}%
%     \label{appendix:\Alph{AppendixCounter}} % Automatic label based on appendix letter
% }

% \setcounter{figure}{0}
% \setcounter{table}{0}
% \renewcommand{\thefigure}{\theAppendixCounter\arabic{figure}}
% \renewcommand{\thetable}{\theAppendixCounter\arabic{table}}

% \input{Appendix}

% \end{appendices}

\end{document}

%% file: intro.tex
The rapid development of Artificial Intelligence (AI) technologies has made it possible to automate decision making in many  domains. However, it has been discovered that 
AI models often acquire pre-existing biases in the dataset used for their training, resulting in the unfair treatment to individuals from different demographic backgrounds~\cite{mesa_2021,dastin_2018}.  
%including criminal justice~\cite{mesa_2021} and recruitment~\cite{dastin_2018}. 
This increased awareness of fairness issues of AI has led to many recent studies in understanding people's fairness perceptions of and reactions to AI models~\cite{haiyiPaper,altugPaper,harrison2020empirical}. 
These studies look into the perspectives of different stakeholders, among which a key stakeholder is {\em decision subjects}, the people who are actually subject to the AI models' decisions. For example, \citeauthor{haiyiPaper}~\shortcite{haiyiPaper} found that when decision subjects only interacted with an AI model once, both the model's unbiased decisions across different groups of subjects 
and the model's favorable decisions towards subjects' own group resulted in an increase in the perceived fairness of the AI model.

Meanwhile, %a recent line of literature 
recent research 
on the long-term dynamics and implications of fairness in AI~\cite{liu2018delayed,liu2020disparate,zhang2020fair,zhang2021fairness,zhang2019group} 
%,d2020fairness,hu2018short,heidari2019long,mouzannar2019fair,puranik2022dynamic} 
has drawn the community's attention to the fact that in reality, decision subjects could often interact with an AI model \textbf{\em repeatedly} over a long term. %During these long-term, repeated interactions with an AI model, 
%Because of the opportunities to interact with an AI model repeatedly, 
Moreover, in each interaction, decision subjects may no longer passively accept the AI model's decisions on them as is. Rather, the AI model's decisions on subjects may shape how they \textbf{\em actively and strategically respond to the AI model}. %, which may
%further impact their perceptions of the model in the long run. 
%\ag{Consequently, these studies have shown the value of studying long-term AI usage to gain a deeper understanding of how end-users perceive AI decisions over time.} 
%\ag{For instance, when an AI model is used to make decisions about individuals (e.g., determine whether to approve one's loan application),
%} 
For instance, one strategic response decision subjects could take in their repeated interactions with the AI model
%\ag{services such as banks that leverages AI decisions,} 
is to decide whether to continue interacting with it 
%the AI model 
and be subject to its  decisions~\cite{zhang2019group,altugPaper}---decision subjects have the freedom to quit using an AI model if they wish so. 
%\ag{considering the commercial settings where} 
As another example, decision subjects could also respond to AI decisions on them by investing in effort to 
%change their ``profile''
improve themselves, hoping for an increased chance of receiving the favorable decision in the future~\cite{zhang2020fair,liu2020disparate}---job applicants could take additional courses about a skill, and loan applicants could explore options to increase their credit scores, both aiming to improve their ``{\em qualification}'' for the favorable decision (i.e., getting the job offer or the loan approval). %In fact, the possibility for decision subjects to respond to AI decisions on them by improving their qualifications 
The possibility to improve one's qualification %naturally requires 
may encourage
decision subjects to take a long-term view to think about the future when responding to the AI model at present, 
%as a response to AI decisions on them, decision subjects' attempts to improve their qualifications 
and it may %not only 
indirectly influence their willingness to continue interacting with the AI model.
%but also make it possible for the deployed AI model to change the characteristics of underlying subject populations over time. %Such change may further feed back into the AI model's decisions~\cite{liu2020disparate,d2020fairness}, and invite questions on the impacts of AI-based decisions on the long-term welfare and well-being of different groups of decision subjects (e.g., will AI decisions lead to  eventual equality between the qualifications of various groups? \cite{zhang2020fair,mouzannar2019fair}).

As decision subjects could  repeatedly and strategically respond to AI decisions in many real-world scenarios, 
%the questions of 
understanding their reactions to AI models with different fairness properties and what they perceive as ``fair'' become critical again.
%Thus, to better understand \ag{decision subjects'} fairness perceptions of and reactions to AI models in more realistic settings, %these more realistic repeated interactions, 
%one may naturally 
Specifically, we ask that when decision subjects can strategically and repeatedly respond to AI decisions: 
\squishlist
    \item \textbf{RQ1}: How will the AI model's fairness properties (both across groups and on the subject's group) affect decision subjects' engagement with the model (e.g., willingness to improve themselves and to be subject to AI decisions)?
    \item \textbf{RQ2}: How will the AI model's fairness properties affect decision subjects' perceived fairness of the model?
\squishend

Predicting answers to these questions turns out to be very challenging. In terms of the willingness to improve one's qualification for the favorable decision, it is possible that decision subjects decide how much to improve themselves %their own qualification 
solely based on whether doing so increases their utility, 
%the gap between their ``desired'' qualification level and their current qualification level, 
and is not affected by the AI model's decision fairness at all. However, one may also speculate that 
%about the influence of an AI model's decision fairness on a subject's motivation to improve their qualifications. Specifically, 
the AI model's biases against 
a subject's group could %substantially 
affect their drive to improve themselves. For example, if the AI model consistently places a subject's group at a disadvantage position in its decisions, 
%when making favorable decisions, 
individuals in that group might feel they are being treated as ``second-class citizens''. 
This feeling could diminish their motivation to improve their qualifications. On the other hand, recognizing the bias, they might be even more determined to improve, seeing it as the sole avenue to increase their odds of favorable decisions and level the playing field with people from other groups. Without a clear hypothesis on how the AI model's decision fairness affects decision subjects' willingness to improve their qualification, predicting how their willingness to keep interacting with the AI model or their perceived fairness of the AI model are affected by the AI model's fairness properties also becomes difficult. This is because both retention and fairness perceptions can be highly influenced by the final qualification level that decision subjects could reach. 

To complicate things further, answers to these questions may also vary across different contexts. For example, one may conjecture that if improving one's chance of getting the favorable decision is particularly difficult for those who really ``need'' the improvement (i.e., those with relatively low qualification), the impact of AI fairness on decision subjects may be more salient, as the hope of changing one's fate through efforts and self-improvement is limited. Similarly, it is also possible that the impact of AI fairness on decision subjects is larger if AI exhibits discriminatory behavior on some salient protected social attributes, triggering people's strong emotional attachment to their own group identities. Formally, one may ask that when decision subjects can strategically and repeatedly respond to AI decisions:

\squishlist
\item \textbf{RQ3}: Do answers to RQ1--RQ2 change if the difficulty for decision subjects to improve their qualification vary with their current qualification level in different ways?
\item \textbf{RQ4}: Do answers to RQ1--RQ2 change when the AI model's fairness properties is/is not discussed with respect to groups defined by protected social attributes, such as gender?
\squishend

To answer these questions, we conducted three exploratory human-subject studies on Amazon Mechanical Turk ($N=368$, $713$, $416$ for the three studies, respectively). In all three studies, subjects 
%in our experiments 
completed a simulated loan application task that was %carefully 
designed to mirror real-world loan application scenarios where the loan decisions are made by an AI model. %As the subject interacted with the AI model by applying for loans from it, 
Subjects were free to decide how many times to apply for a loan from the AI model, and whether to improve their own qualification (i.e., their credit score) before each application. 
In each study, we created two treatments to reflect that the AI model may or may not show systematic bias towards one group over the other in granting loans.
%varying levels of fairness in the AI model's decisions---in the ``fair AI'' treatment, the AI model made unbiased decisions on subjects of different groups, 
%in red and blue groups, 
%while in the ``unfair AI'' treatment, the AI model systematically favored subjects from one group over the other %from the red group over those from the blue group 
%in granting loans. 
Study 1 was designed to answer \textbf{RQ1} and \textbf{RQ2}, so the difficulty for subjects to improve their qualification does not vary with their current qualification level, and subject's group identity was randomly assigned. To answer \textbf{RQ3}, we slightly varied the design of Study 2 so that the difficulty of qualification improvement either increased or decreased with the subject's current qualification level. Finally, to answer \textbf{RQ4}, subject's group identity in Study 3 was decided by their self-reported gender rather than a randomly assigned value.
%Therefore, Study 1 enabled us to examine the impacts of AI fairness on decision subjects' engagement with and perceived fairness of the AI model as they strategically respond to the AI model's decisions on them, when the difficulty of qualification improvement does not vary with one's current qualification level and the fairness of the AI model is not examined with respect to salient protected attributes. 
%To understand the generalizability of our findings, we conducted two more replication studies. The only difference between Study 2 ($N=713$) and Study 1 was that in Study 2, we varied the difficulty of qualification improvement to either increase or decrease with the subject's current qualification level. Moreover, Study 3 ($N=416$) was the same as Study 1, except for subjects' group identity was set to be their self-reported gender rather than a randomly assigned value, and correspondingly, the fairness properties of the AI model was examined with respect to gender. 

Our results show that when decision subjects could repeatedly and strategically respond to the AI model's decisions on them,  
%in their repeated interactions with the model, 
their engagement with the model---including their willingness to improve their qualification and willingness to keep interacting with the model---are {\em not} influenced by the AI model's decision fairness. 
This holds true both when the difficulty of qualification improvement changes with one's current qualification level in different ways, and when the AI model's fairness is/is not examined with respect to salient protected attributes like gender.
However,  
we find that 
despite the possibility of strategic responses, 
decision subjects still perceive the AI model as less fair if the model biases against the subjects' group by placing them at a disadvantaged position in receiving the favorable decision. 
%\erase{The AI model's decision fairness also has a larger impact on decision subjects' perceived fairness of the model when it is more difficult for people with low qualification to improve their chance of getting the favorable decisions. In sum, when decision subjects can strategically and repeatedly respond to AI decisions, their level of engagement with different AI models does {\em not} necessarily match their fairness preferences among them. }
In other words, when decision subjects can strategically and repeatedly respond to AI decisions, their level of engagement with an AI model does {\em not} reflect either the model's group-level decision fairness or their perceived fairness of the model. %their fairness preferences among them. 
We conclude by providing possible explanations for our findings and discussing their implications, %limitations, and future work.

%% file: related.tex
%An increasing line of work has focused on defining, understanding, and enforcing fairness in AI in recent years. For example, many studies proposed different ways of defining fairness and formalized these definitions as algorithmic constraints~\cite{DBLP:journals/corr/abs-1908-09635,10.1145/3194770.3194776,AlFr,gajane2018formalizing,dwork2011fairness,NIPS2016_9d268236,10.1093/qje/qjx032}.
%However, it has been recognized that there is no universal definition of fairness that can be applicable for all kinds of contexts and satisfies diverse expectations and requirements~\cite{universalFairness,srivastava2019mathematical,contextFairness}. 

The complexity of the notion  of ``AI fairness'' %, \erase{together with the advocate for better understanding of real human users' perceptions and usage of AI~\cite{liao2021human,lai2021towards,poursabzi2021manipulating},} 
has inspired many studies on understanding whether and when do humans perceive an AI model as ``fair'' in decision making~\cite{yaghini2021human, 10.1145/3306618.3314248,harrison2020empirical,srivastava2019mathematical}. 
Of particular relevance to our study %in this paper 
are a few recent works on understanding {\em decision subjects}' fairness perceptions of an AI model that makes decisions about them. Earlier studies typically focus on one-shot interaction scenarios where decision subjects would only receive a decision from an AI model once. For example,  \citeauthor{yurrita2023disentangling}~\shortcite{yurrita2023disentangling}  studied the influence of explanations, human oversight, and contestability on fairness perceptions of the decision subjects in a one-shot interaction with the AI model for loan approvals. They found that while explanations and contestability significantly impacted fairness perceptions, human oversight showed minimal effect. In another study involving one-shot interaction with the AI model, decision subjects perceive an AI model as fairer both if the AI model makes a favorable decision on them and if the AI model is not biased towards or against any particular group~\cite{haiyiPaper}.

More recently, there is a line of theoretical works on ``long-term fairness''~\cite{liu2018delayed,hu2018short,zhang2019group,d2020fairness,mouzannar2019fair,heidari2019long} emphasizing that in the real world, decision subjects often engage in {\em repeated} interactions with the AI model, and the dynamics between the AI model's decisions on subjects and subjects' {\em strategic} reactions to those decisions could create feedback loops. One %real-world 
domain that is frequently studied in the long-term fairness literature is loan lending---For a loan applicant characterized by a profile $\mathbf{x}$, the bank may use its AI-based loan approval system to make loan lending decisions on the applicant. Moreover, the applicant may strategically respond to these decisions by staying or leaving the system and/or changing their profile $\mathbf{x}$, 
which may further change the AI model's training data and impact the model's decision-making policy in the future. This %advocate for moving beyond 
shift from one-shot to long-term, repeated interaction has inspired some empirical studies looking into 
%raises the importance of understanding 
how do decision subjects react to and perceive AI models in 
%\ag{the latter setting.}  %
their long-term interactions with AI. 
For example, it was found that when decision subjects %could %repeatedly interact with an AI-based decision system and 
had the choice to leave the AI-based decision system at any time in their repeated interactions, their willingness to stay in the system and their perceived fairness of the system are 
%not significantly affected by the AI system's unbiased treatment across groups, but mostly determined by 
significantly affected by whether the system
is in favor of the subject's own group~\cite{altugPaper}. 
Compared to earlier works, in this study, we take into account another key strategic action that decision subjects could take in their repeated interactions with AI.  
That is, decision subjects could also freely decide whether to improve their qualifications for receiving future favorable decisions from AI. %themselves to increase their chance of getting the favorable decision from AI in the future. 
In the real world, these qualification improvement attempts are usually realized through the adjustment of decision subjects' input attributes, possibly as the decision subjects follow the algorithmic recourse plans suggested by the AI model to change their situation towards receiving a more favorable decision%favorable outcomes from the AI-based decisions
~\cite{ustun2019actionable}.

%\erase{
%There are a few reasons for us to 
We believe that the addition of qualification improvement in the set of decision subjects' strategic actions 
%that decision subjects may take in their repeated interactions with the AI model 
brings new perspectives for re-examining the relationship between AI's decision fairness and decision subjects' reactions to and perceptions of it.  
%in more realistic scenarios. 
Indeed, the possibility to improve their qualification for future favorable decisions may shift decision subjects' attention from {\em focusing on the present} to {\em thinking about the (long-term) future}, which %Through making the future seem more connected to the present, this shift 
may have complicated implications on how decision subjects would react to the AI model at present. For example, because of their future thinking, decision subjects may change how much risks they are willing to take or how they treat the immediate and future rewards~\cite{thorstad2018big,hershfield2011future}. It may also change the ways that decision subjects weigh utility-related considerations (e.g., how many favorable decisions can I get from the AI model in the long run?) and fairness-related considerations (e.g., is the AI model's decision on me fair?) in deciding their engagement with the AI model. It is even possible for decision subjects to change how they define ``fairness'' for AI. This is because the possibility of qualification improvement may allow decision subjects to evaluate the AI fairness not only through ``{\em social comparison}'' (e.g., does AI make similar decisions on my group and other groups? \citeauthor{festinger1954theory}~\shortcite{festinger1954theory}) but also through ``{\em temporal self-comparison}'' (e.g., does AI grant more favorable decisions to me after my qualification is improved?  \citeauthor{albert1977temporal}~\shortcite{albert1977temporal}). 
A few theoretical works have examined the implications of AI fairness in scenarios where decision subjects can respond to AI decisions 
strategically by improving their qualifications
%repeatedly and strategically%found that ensuring fairness constraints in an AI model in a one-shot setting does not necessarily promote the equality in the qualification rates of different groups in the long run
\cite{zhang2020fair,liu2020disparate,mouzannar2019fair}. However, none of these works focus on {\em empirically} characterizing how decision subjects would respond to AI models with different levels of fairness to determine their qualification transitions. 
Rather, they make simplified 
assumptions about subjects' behavior in their theoretical derivations. %\new{~\cite{mouzannar2019fair,liu2020disparate,zhang2020fair}}. %\erase{For example, \citeauthor{mouzannar2019fair}~\shortcite{mouzannar2019fair} assumed that the chance for a decision subject to improve their qualification is only decided by how likely subjects with different qualification levels in {\em their own group} receive the favorable decision from AI, i.e., it is not influenced by the intra-group disparity in the AI decisions. }
%AI model's decisions on subjects of other groups at all. 
%Similarly, 
For example, \citeauthor{liu2020disparate}~\shortcite{liu2020disparate} assumed that decision subjects decide about their qualification improvement based on a cost-benefit analysis by examining whether the increase in utility after the qualification is improved is larger than the cost of improvement. %, and is not affected by the intra-group disparity in the AI model's decisions. 
\citeauthor{zhang2020fair}~\shortcite{zhang2020fair} considered whether decision subjects received a favorable decision from the AI model in one interaction as the key influencing factor for them to decide whether to improve their qualification in the next interaction. %, while the specific influence can be dependent on the subject's group. 
Different from these theoretical works, we use experiment with real human subjects 
%having human subjects actually interact with different AI models in our experiments, 
to provide empirical insights into whether and how the AI model's decision fairness affects decision subjects' 
qualification transitions in repeated interactions, and their engagement with/perceived fairness of the AI model in general.  
Thus, results of our study may help verify or reject assumptions made in previous works.

%% file: study1.tex
To understand how an AI model's decision fairness affects decision subjects' repeated and strategic interactions with the AI model, 
%decision subjects' long-term interactions with AI models 
%when they could strategically respond to the AI decisions by voluntarily determining whether to improve their qualification for the favorable decision and whether to continue interacting with the AI model,  
we conducted a series of human-subject experiments\footnote{All of our experiments were approved by the IRB of the authors' institution.}. In Study 1, we aim to first answer \textbf{RQ1}--\textbf{RQ2} in an environment where (1) the AI model's fairness properties are {\em not} examined with respect to a salient protected attribute, and (2) the difficulty for a decision subject to improve their qualification for the favorable decision does {\em not} change with the subject's current qualification level.

\subsection{Experimental Design}
\subsubsection{Tasks.}
\label{sec:task}

Subjects in our experiment were asked to complete a simulated loan application task, which was carefully designed to mimic the real-world loan application scenario that is often used in theoretical studies of the long-term dynamics of AI fairness~\cite{liu2018delayed,zhang2020fair,d2020fairness} and experimental studies of fairness perceptions of AI~\cite{altugPaper,yurrita2023disentangling}.  
%in our experiment. 
%as that used 
%in
%~\citeauthor{altugPaper}~\shortcite{altugPaper}. 
Specifically, each subject was assigned with a randomly-generated persona of a small business owner containing 5 attributes; this persona was used as the subject's loan application ``profile''. We highlight two key attributes in the persona: %\ag{We adopted two key features for each subject's persona from the long-term fairness literature that extensively used them to model loan lending pipeline and here we highlight only those two features:}  

\squishlist
\item \textbf{Group identity}: The subject's group membership, which was set to be either ``red'' or ``blue''. This is to reflect that in Study 1, the AI model's fairness properties across groups are {\em not} defined on salient protected attributes.
\item \textbf{Initial credit score level}: The subject's credit score level at the beginning of the experiment, which was taken from 
%one of the 9 possible ranges in 
the set %of 
\{300--350, $\cdots$, 500--550, $\cdots$, 700--750\}; the subject could later decide to ``improve'' their credit score with some cost (see details below). 
%We told subjects that their credit score level largely reflects their ``qualification'' for the loans---credit scores of 650 or higher were generally considered as ``high'', and higher credit scores were associated with higher chance of getting loans approved.

\squishend

%We set a subject's group identity as one of the two arbitrary values (red vs. blue) to   
%(in Study 3, we will examine the generalizability of our results in settings where the AI model's fairness properties are defined with respect to salient protected attributes). 
In addition to the above two attributes, the subject's persona also included three other attributes---their number of years of having a credit history, their home ownership status (e.g., rent or own), and the type of small business they ran (e.g., healthcare, construction). For each attribute in a subject's persona, we uniformly randomly sampled a value from the set of all candidate values for that attribute. 
After getting their assigned profile, the subject was asked to use it to apply for loans from their ``local bank'' to support their business, and they were explicitly told that the bank utilized {\em an AI model} to analyze loan application profiles and make loan approval decisions. %approve or deny applications. 
The subject was also informed that %the bank's AI model would 
%analyze the subject's loan application profiles %(i.e., the subject's persona in the experiment) 
%to make a loan approval decision, while the subject's 
their credit score level would be used by the AI model as a primary determinant of their ``qualification'' for the loans---credit scores of 650 or higher were generally considered as ``high'', and higher credit scores were associated with higher chance of getting loans approved. 
%while other aspects of their persona would be taken into account for loan approval, their credit score level would be the primary determinant of a favorable decision
%\footnote{\ag{In reality, the AI model determined loan approvals based solely on the subjects' group identity and their most recent credit score level.}. They were advised to concentrate on this attribute, as higher credit score levels were associated with a greater likelihood of loan approval}

\begin{figure*}[t!]
    \centering   \includegraphics[width=\textwidth]{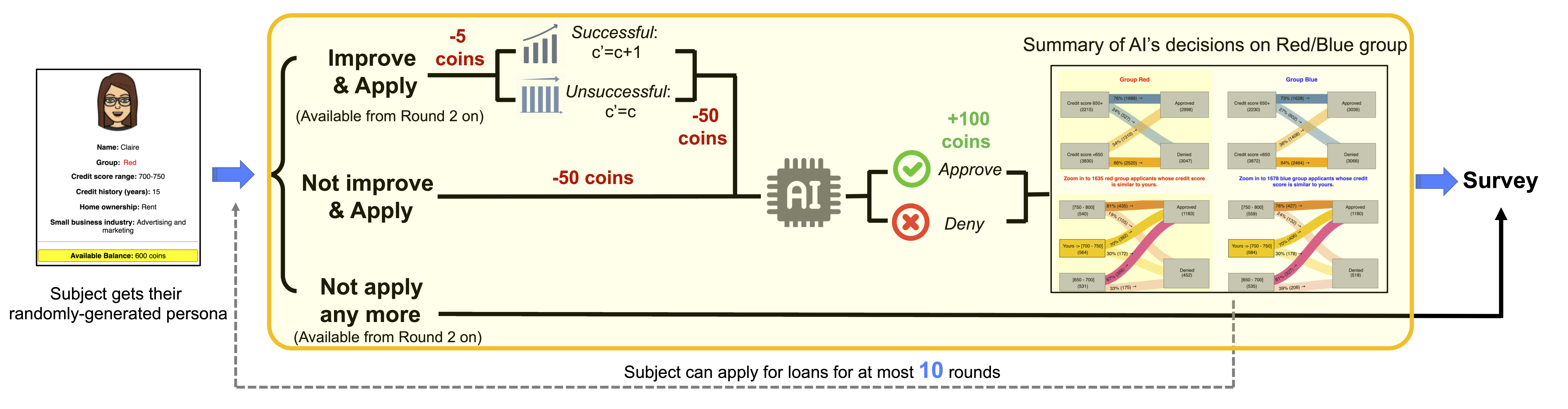}
   \vspace{-18pt}
    \caption{An illustration of the process of the loan application task. Here, $c$ is the subject's current qualification level, while $c'$ denotes the qualification level after an improvement attempt, which can either remain the same or advance to the next level.}~\label{fig:task}
   \vspace{-10pt}
\end{figure*}

In the experiment, the subject started the experiment with 600 ``coins'' in their account. 
%, and they could interact with the bank's AI model by applying loans from the bank. 
%, and they could apply for loans from the bank for {\em at most} 10 rounds. 
Each loan application cost the subject 50 coins, and the subject would gain 100 coins if the application got approved or nothing otherwise. %\ag{\footnote{\ag{The coin parameter amounts used were those of~\citeauthor{altugPaper}.}}} 
Each subject was asked to apply for a loan from the bank for at least once to get a sense of the AI model's decision fairness. After that, they could interact with the AI model for at most 9 more rounds. In each round, 
the subject had the freedom to choose from one of three actions:
\squishlist
\item \textbf{Improve and apply}: The subject would first attempt to improve their credit score to the next level (e.g., from 600--650 to 650--700) with a cost of 5 coins\footnote{
%We set the cost of improving one's qualification to be 5 coins, because 
%We used a Markov Decision Process to simulate a rational subject's decision making process and 
Via a simulation study, we found that an improvement cost of 5 coins is an intermediate level of cost that requires subjects to carefully deliberate about whether and when to improve their qualification. If the improvement cost was too low (or high), subjects may simply opt for always improving their qualification (or never improving their qualification). See Appendix~\ref{sec:simulation}  for more details.  
}, and then apply for a loan with a cost of 50 coins. The credit score improvement attempt was {\em not} guaranteed to be successful (see details below)---if successful, subjects would be notified, %and the AI model's loan approval decision on the subject would be based on the updated credit score level
and the AI model would use the updated credit score to make the loan approval decision. 
%\footnote{The highest credit score level a subject could reach was 800--850. Once reaching this level, subjects could not further improve their credit score.}.
\item \textbf{Apply without improvement}: The subject would directly apply for a loan with a cost of 50 coins without attempting to improve their credit score level. 
This action was provided to subjects to reflect that in reality, whether and when to improve one's qualification is a voluntary (and strategic) decision. 
%Given that improving one's qualification in the real world is both voluntary and can incur expenses, subjects were given the option to apply for a loan without necessarily first trying to improve their credit score.
\item \textbf{Not apply any more}: The subject would not apply for loans any more and would be redirected to the end of the experiment. This action was provided to subjects to reflect that in reality, 
%subjects have the freedom to decide 
whether to continue interacting with an AI model is a voluntary (and strategic) decision. 
\squishend

Since Study 1 concerns an environment where the difficulty for decision subjects to improve their qualification does {\em not} change with their current qualification level, we set the ``success rate'' for subjects across \emph{all} credit levels to progress to the next level at a constant value (i.e., 44\%). %(i.e., 44\%; we will vary how the success rate changes with subjects' current qualification level in Study 2 to examine the generalizability of our results). %\footnote{In the other two improvement schemes where the difficulty for decision subjects to improve their qualification {\em does} change with their current qualification level, the average success rate across subjects of different credit levels is 44\%. Detailed explanations of these schemes will be provided in subsequent sections.}. 
%, should they decide to improve their qualification. 
That is, if the subject attempted to improve their qualification in one round, whether they could successfully progress to the next credit level would be stochastically decided by this success rate\footnote{Subjects were not explicitly told about this success rate but could experience it through their actual improvement attempts.}. Once the subject successfully progressed to the next credit level, they would at least maintain that level, and possibly progress to even higher levels if they decided to make additional improvement attempts in future rounds.

At the end of each round, 
if the subject applied for a loan in that round, %(i.e., they either ``improve and apply'' or ``apply without improvement''), 
the AI model's approval or denial decision on the subject would be revealed to them, without providing explanations on why it makes this decision. %(see Section ``Experimental Treatments'' for details on how the AI model makes its decisions). 
To enable subjects to perceive the AI model's decision fairness, a flowchart summary of the AI model's decisions in this round on applicants of both red and blue groups %groups, especially for those with similar credit scores as the subject, %(i.e., applicants with the same credit score level as the subject or one level below or above the subject), 
would also be provided to the subject (see Figure~\ref{fig:flowcharts} in appendix for an example). 
%; more details are discussed below in ``Treatments''%Section~\ref{sec:treatments}
%). %the "Treatments" subsection for details and refer to the flowcharts in 
%). %\my{Maybe include the screenshots for the flowchat of fair AI vs. unfair AI as an example.} \ag{referred.}

Figure~\ref{fig:task} illustrates the process of the simulated loan application task. We note that the designs of this task reflect a few key characteristics of the real-world repeated interactions between decision subjects and an AI model: (1) participating in the decision making process (thus triggering the usage of the AI model) is costly; (2) receiving a favorable decision from the AI model is rewarding; (3) the improvement of qualification is costly and has uncertainty; and (4) decision subjects have the freedom to respond to the AI model's decisions by deciding whether to improve their qualification and/or whether to continue being subject to the AI model's decision. By assigning each subject with a persona of a small business owner, our experiment reflects a scenario where the real-world decision subjects will often interact with the AI-based decision systems {\em repeatedly}, as small business owners often need to apply for loans at different time points to meet their business's various financing needs.
%\footnote{\new{Indeed, small business owners often need to apply for loans at different time points to meet their business's various financing needs, in the U.S., there is no limit on the number of Small Business Administration (SBA) loans\footnote{\url{https://www.sba.gov/funding-programs/loans}.} that businesses can apply for, and it is found that it takes six months or less for most business owners to spend their funds from the most recent loans~\cite{forbes}. 
%Our experimental task abstracts the long-term, repeated interactions between decision subjects and an AI model in a real-world loan lending context, albeit the interactions are condensed into a shorter timeframe of 10 rounds in the task.}}. 

%\ag{Our web application was developed using the Meteor platform\footnote{https://www.meteor.com/}, which could accommodate 100 subjects participating in our task simultaneously. We utilized standard HTML for the user interface components and intro.js\footnote{https://introjs.com/} to guide subjects through the different components of the interface. Additionally, we employed plain JavaScript to create forms and animations that brought the interactive elements of the web-app to life.} 

\begin{table*}[t]
\centering
\footnotesize % smaller font size
\renewcommand{\arraystretch}{1.2} % adjust vertical spacing
\begin{subtable}[t]{0.2\textwidth} % reduced width for each subtable
\resizebox{\linewidth}{!}{ % adjust table size
\begin{tabular}{ | P{6em} | P{1cm}| P{0.5cm} | } % further reduced column widths
  \hline
  Credit/Decision & Approve & Deny \\ 
  \hline
  $800$--$850$ & 85\% & 15\% \\ 
  \hline
  $750$--$800$ & 78\% & 22\% \\ 
  \hline
  $\vdots$ & $\vdots$ & $\vdots$ \\
  \hline
  $350$--$400$ & 22\% & 78\% \\ 
  \hline
  $300$--$350$ & 15\% & 85\% \\ 
  \hline
\end{tabular}
}
\caption{Fair AI: Red/Blue group}~\label{tab:fair}
\end{subtable}
\hspace{0.5em}
\begin{subtable}[t]{0.2\textwidth} % reduced width for each subtable
\resizebox{\linewidth}{!}{ % adjust table size
\begin{tabular}{ | P{6em} | P{1cm}| P{0.5cm} | } % further reduced column widths
  \hline
  Credit/Decision & Approve & Deny \\ 
  \hline
  $800$--$850$ & 95\% & 5\% \\ 
  \hline
  $750$--$800$ & 88\% & 12\% \\ 
  \hline
  $\vdots$ & $\vdots$ & $\vdots$ \\
  \hline
  $350$--$400$ & 32\% & 68\% \\ 
  \hline
  $300$--$350$ & 25\% & 75\% \\ 
  \hline
\end{tabular}
}
\caption{Unfair AI: Red group}~\label{tab:unfair:red}
\end{subtable}
\hspace{0.5em}
\begin{subtable}[t]{0.2\textwidth} % reduced width for each subtable
\resizebox{\linewidth}{!}{ % adjust table size
\begin{tabular}{ | P{6em} | P{1cm}| P{0.5cm} | } % further reduced column widths
  \hline
  Credit/Decision & Approve & Deny  \\ 
 \hline
  $800$--$850$ & 75\% & 25\% \\ 
  \hline
  $750$--$800$ & 68\% & 32\% \\ 
  \hline
  $\vdots$ & $\vdots$ & $\vdots$ \\
  \hline
  $350$--$400$ & 12\% & 88\% \\ 
  \hline
  $300$--$350$ & 5\% & 95\% \\ 
  \hline
\end{tabular}
}
\caption{Unfair AI: Blue group}~\label{tab:unfair:blue}
\end{subtable}
\vspace{-10pt}
\caption{The AI model's probability of approving/rejecting loan applications in different treatments.}
\label{tab:conf_mat_exp1}
\vspace{-18pt}
\end{table*}

\subsubsection{Treatments.}
\label{sec:treatments}

%By varying the fairness level of the bank's AI model across groups, i.e., whether the AI model exhibited systematic biases towards applicants from a certain group, 
We created two treatments in Study 1: 

\squishlist
\item \textbf{Fair AI  model}: In this treatment, the bank's AI model was fair towards subjects in both the red group and the blue group. 
%adopts a fair treatment approach towards both groups and employs a \emph{single} "decision matrix"---as demonstrated in Table~\ref{tab:fair}---in other words \emph{without the group membership of the applicants} to arrive at a stochastic lending decision for both of the groups. 
In particular, as shown in  Table~\ref{tab:fair}, {\em regardless of the subject's group identity}, the AI model's loan approval rate always started from 15\% for subjects with the lowest credit level (i.e., 300--350), and the approval rate increased by 7\% as the subject's credit score went one level up, with a highest possible approval rate of 85\% for subjects with the highest credit level (i.e., 800--850). 
%with the approval rate for the highest range being 85\% and the lowest having a rate of 15\%.
\item \textbf{Unfair AI model}: In this treatment, the bank's AI model was unfair and {\em systematically biased against subjects in the blue group}. 
Table~\ref{tab:unfair:red} shows this model's approval rate for subjects in the red group, while Table~\ref{tab:unfair:blue} shows this model's approval rate for subjects in the blue group. As shown in the tables, for every credit level, the AI model's approval rate for red group subjects with this credit level was always 20\% higher than that for blue group subjects\footnote{%Compared to that in the ``fair AI'' treatment, given any credit level, in the ``unfair AI'' treatment, the approval rate for subjects of the red group increased by 10\% while the approval rate decreased by 10\% for subjects of the blue group. 
Note that since the subject's group identity was sampled uniformly randomly from the two candidate values,  
%(i.e., red vs. blue), 
for each credit level, the 
expected chance for a subject getting their loans approved in the ``unfair AI'' treatment was still the same as a subject with the same credit level in the ``fair AI'' treatment.}. Same as that in the previous treatment, the AI model's increment in approval rate for each increased credit level was kept at 7\% for both red and blue groups. 
\squishend

So, if the subject decided to apply for a loan in one round, the AI model's loan approval decision on them would be stochastically decided by the approval rate for their most updated credit level, given the subject's treatment assignment and group identity. As described earlier, at the end of each round where the subject applied for a loan, we also presented to the subject a summary of the AI model's decisions across applicants in different groups. In particular, we told the subject that there are a total of $N\sim U[12000, 12200]$ applicants who applied for loans from the bank in this round. 
%, where the value of $N$ was uniformly randomly drawn from the interval of $[12000, 12200]$. 
For each of these $N$ applicants, we randomly generated their persona, and then simulated the AI model's decision on them using the approval rates defined by {\em the subject's assigned treatment}. We then visualized
the AI model's decisions on all $N$ applicants using two sets of flowcharts. The first set showed the AI model's approve/deny decisions for applicants with/without ``high'' credit scores (i.e., a score of at least 650). The second set zoomed in to applicants with similar credit scores as the subject and showed the AI model's approve/deny decisions for applicants whose credit level was one level higher than, the same as, or one level lower than the subject (see  Figure~\ref{fig:flowcharts} in appendix)\footnote{We used flowcharts to visualize the AI model's decisions because prior research has showed their effectiveness in aiding the comprehension of algorithmic model performance among non-experts~\cite{shen2020designing}. %To ensure subjects correctly understand the information included in the flowcharts, 
We also provided textual explanations to help subjects interpret the numbers shown in the flowcharts.  %(e.g., ``For the 2214 red group applicants whose credit score is 650 or higher, 1866 (84\%) of them get their applications approved.'').
}. Note that within each set, the AI model's decisions on red/blue group applicants were shown in separate flowcharts; this enables subjects to make direct comparisons across the two groups of applicants to determine the AI model's decision fairness. 
%in separate flowcharts.

\subsubsection{Procedure.}
\label{sec:procedure}
Our experiment was made available as a Human Intelligence Task (HIT) on Amazon Mechanical Turk (MTurk). Only U.S. workers with a HIT approval rate of at least 95\% and a total of at least 1000 approved HITs were eligible for taking this HIT. 

Upon arrival, the subject first created a nickname and selected an avatar for their persona. Next, we provided an interactive tutorial explaining each attribute's meaning on the subject's randomly-assigned persona profile. Additionally, this tutorial explained to subjects what they were asked to do in each round, how to use the interface, 
%in order to utilize the AI model for receiving a decision (e.g., clicking on an interactive button that simulates the AI model's decision-making process), 
and how to interpret the summary information of the AI model's decisions as displayed in the flowcharts. 
%The AI model's decisions, as detailed in subsection~\ref{sec:treatments}, encompassed the AI's decisions on simulated applicants from both the subjects' own group and the other group. Thus, the decisions concerning themselves and others constituted the key information for them to understand the AI model's fairness toward all applicants.} 
%To assess subjects' understanding of the tutorial, 
We then used a quiz of 5 questions to test subjects' understanding of the task procedure and ability to interpret flowcharts. Subjects were only allowed to advance to the actual experiment after correctly answering all 5 questions.

In the actual experiment, the subject was first randomly assigned to one of the two treatments.  
%as described in Section~\ref{sec:treatments}. 
%the ``Treatments'' subsection. 
Then, the subject went through the simulated loan application task. Note that throughout the task, the subject's credit level would be updated if the subject's improvement attempt was successful, and the subject's account balance would be updated depending on the loan application outcomes. 
Once the subject finished the loan application task, they were redirected to our post-experiment survey. The survey included questions about the subject's demographics (e.g., race, age, education) and perceived fairness of the AI model. We also measured a few other characteristics of the subject that we conjectured to influence their behavior in interacting with the AI model, such as their risk attitude.  %(see %\old{Section~\ref{sec:cov}} 
%Section~\ref{sec:analysisMethods} for more 
%details below). 
All questions except for the demographics were presented as 5-point Likert scale questions in which the subject needed to indicate their agreement with a set of statements from 1 (strongly disagree) to 5 (strongly agree). The subject could also explain why they considered the AI model they encountered in the experiment as fair or unfair using free-form texts. The complete list of survey questions are provided in Appendix~\ref{sec:survey}. %\hyperref[sec:appendix]{appendix}\footnote
%{\new{See \hyperref[sec:interface]{appendix} for our checks to determine the integrity of the data collected.}}.
%and questions to measure our variables of interest (see Section~\ref{sec:Operationalization}).
% perceived fairness of the AI model, sensitivity to fairness, risk attitude, and level of empathy. All questions except for the demographics were presented as 5-point Likert scale questions in which the subject needed to indicate their agreement with a series of statements from 1 (strongly disagree) to 5 (strongly agree). Specifically, statements regarding subjects' perceived fairness of the AI model (e.g., ``The bank’s AI system is fair to manage loan applications.'') were adapted from~\cite{haiyiPaper}. Statements on subjects' sensitivity to fairness (i.e., how much the subject holds fairness as a core value; an example statement was ``I would stop using an AI system if it is unfair, even if it tends to be in favor of me'') were adapted from~\cite{altugPaper}. Finally, statements related to the subject's risk attitude (e.g., ``I like to do frightening things.'') and level of empathy\footnote{We included the level of empathy that an individual exhibits as a covariate in our analysis, since highly empathetic people may be more concerned about the fairness of an AI model, even if the AI model is not biased against them. } (e.g., ``I get a strong urge to help when I see someone who is upset.'') were taken from \cite{riskAtti} and~\cite{spreng2009toronto}, respectively. A detailed list of the survey questions can be found in the supplemental materials. 

Subjects were told that their payment in this experiment was composed of a {\em base payment} of \$2 and a {\em bonus payment} that would be decided by their account balance at the end of the experiment. Upon survey completion, we converted the remaining balance in the subject's account to their bonus payment using a 500 coins to \$2 ratio.
%the subject received their bonus that was proportional to their remaining account balance (i.e., we converted every 500 coins to \$2.00). The base payment of our experiment was \$2, and 
The maximum bonus a subject could earn was \$4.40. Subjects spent a median time of 27 minutes on our experiment and received a median payment of \$4.30, resulting in an hourly wage of \$9.60. We also included three filtering procedures to filter out potential spammers (see Appendix~\ref{sec:filtering} for details). A subject's data was only considered valid if they passed all filtering procedures.

\subsection{Analysis Methods}
\label{sec:analysisMethods}

%Utilizing subjects' behavior in our experimental task and their responses in the post-experiment survey, 
We used regression analyses to answer our research questions. %\footnote{\new{See \hyperref[sec:appendix]{appendix} Table~\ref{tab:variables} for a summary of the independent variables, dependent variables, and covariates we used in the regressions.}}. %\old{Table~\ref{tab:variables} provides a summary of the independent variables, dependent variables, and covariates that we used in the regressions, which we will detail in the following.}
%We operationalized the subjects' responses from the post-experiment survey to define the primary dependent and independent variables, as well as the covariates, for our regression analyses. Further details are provided below. The entire survey questions administered to the subjects can be found in the Supplementary Materials.
%\old{
% \subsubsection{Dependent variables.}
Specifically, we used three dependent variables %in our regressions 
to quantify decision subjects' engagement with an AI model (\textbf{RQ1}) and their fairness perceptions of it (\textbf{RQ2}): 
%, as they repeatedly interacted with the model:
\squishlist
\item \textbf{Engagement--Improvement}: The number of %times that the subject made a 
qualification improvement attempts the subject made 
in the loan application task; a higher value indicates a higher level of willingness for the subject to improve themselves.
\item \textbf{Engagement--Retention}: The number of times that the subject applied for a loan; a higher value indicates a higher level of willingness for the subject to continue being subject to the AI model's decisions. 
\item \textbf{Perceived Fairness}: %This was obtained through summing up 
The subject's rating to six statements (e.g., ``The bank’s AI system is fair to manage loan applications.'') adapted from~\citeauthor{haiyiPaper}~\shortcite{haiyiPaper} regarding their perceptions of the AI model's fairness; %in the post-experiment survey; 
the higher the rating, the fairer the subject found the AI model to be (the max rating is 30). 
%An example statement 
%regarding subjects' perceived fairness of the AI model 
%was ``The bank’s AI system is fair to manage loan applications.''
%The full set of statements on subjects' perceived fairness of AI were adapted from~\cite{haiyiPaper}.
\squishend
The independent variable we included in the regression reflects the AI model's ``{\em fairness properties},'' which was operationalized in two ways. First, 
%For example, 
to explore how subjects' engagement with and perceived fairness of AI vary with the AI model's decision fairness {\em across groups}, 
we used 
%the first independent variable we considered in our study was: (1) A 
a binary variable \textbf{Fair AI} to represent if the AI model treats the two groups of loan applicants in a similar way (i.e., it was set to 1 for subjects in the fair AI treatment and 0 otherwise). 
%the AI model's fairness level across groups. It is set to 1 for subjects in the fair AI model treatment, indicating a fair treatment across groups, and 0 otherwise. 
Second, to understand how subjects' engagement with and perceived fairness of AI are affected by the AI model's decision fairness {\em on the subject's own group}, we used two other binary variables in our regressions---\textbf{Advantaged} and \textbf{Disadvantaged}, representing if the subject's group was favored or disfavored by the AI model. That is, Advantaged (Disadvantaged) was set to 1 only for red (blue) group subjects in the unfair AI treatment. 

%1) A binary variable indicating whether the subject's group was systematically favored by the AI model. It is set to 1 for red group subjects in the unfair AI model treatment, and 0 for all others. 2) A binary variable indicating whether the subject's group was systematically disfavored by the AI model. It is set to 1 for blue group subjects in the unfair AI model treatment, and 0 for all others.
%}

%\old{
% \subsubsection{Covariates.}
% \label{sec:cov}

Finally, to control for the influences on dependent variables beyond those brought up by the independent variables, we considered a few characteristics of the subjects and included them as covariates:

\squishlist

\item \textbf{Initial Credit Score}: The credit score level that was assigned to the subject at the beginning of the experiment. 
%Each subject's initial credit score level was converted into a numerical value between 0 and 11, with larger values representing higher credit levels. 
We conjectured that subjects with higher credit levels  make fewer improvement attempts as there is smaller room of improvement for them. However, they might be more willing to interact with the AI model and even perceive it as fairer, because they were more likely to receive loan approval decisions from the AI model.  

\item \textbf{Fairness Sensitivity}: The degree to which the subject values fairness as a core principle, which was measured in the post-experiment survey through soliciting the subject's opinions on a set of  statements (e.g., ``I would stop using an AI system if it is unfair, even if it tends to be in favor of me'') adapted from~\citeauthor{altugPaper}~\shortcite{altugPaper}. 
%and their potential sensitivity to direct interactions with AI fairness. 
%For example, agreement with the statement ``I would stop using an AI system if it is unfair, even if it tends to be in favor of me'' indicates a higher sensitivity to fairness. The full set of statements regarding subjects' sensitivity to fairness were adapted from~\cite{altugPaper}. 
We %included fairness sensitivity as a covariate as we 
conjectured that subjects' fairness sensitivity 
%of subjects 
may affect how they react to the AI model's decision fairness. 

\item \textbf{Empathy}: The subject's empathy level, which was measured in the post-experiment survey through soliciting the subject's agreement with a set of statements (e.g., ``I get a strong urge to help when I see someone who is upset'') adapted from~\citeauthor{spreng2009toronto}~\shortcite{spreng2009toronto}.  %Empathy was included as a covariate based on 
We conjectured that individuals with higher levels of empathy may exhibit stronger concerns regarding the fairness of an AI model, as they are concerned with the well-being of others even if they are not personally affected by the AI model's bias.

\item \textbf{Risk Attitude}: The subject's risk attitude, %i.e., the likelihood that the subject would engage in risk-seeking behaviors, 
%as a covariate. It 
which was measured through soliciting the subject's opinions on a set of statements (e.g., ``I like to do frightening things.'') adapted from~\citeauthor{riskAtti}~\shortcite{riskAtti}. 
%Since subjects in our experiment need to consistently confront uncertainty, gains, and losses (as they would in the real-world loan lending scenarios), we conjectured that subjects' risk attitude may influence their willingness to engage with the AI model. For example, 
We conjectured that subjects who were more risk-seeking might be more willing to take actions to improve their qualification or continue interacting with the AI model. Previous research has also found that people with higher risk-taking tendencies %less uncertainty avoidance (i.e., more risk-taking tendencies) 
tend to perceive AI systems as fairer compared to those who are less risk-takers~\cite{nakao2022toward}.

\squishend
We fit our experimental data %we collected from our experiment 
into 
%two sets of 
regression models 
%(i.e., one set using ``Fair AI'' as the independent variable and another set using ``Advantaged'' and ``Disadvantaged'' as the independent variables) 
to predict subjects' engagement (\textbf{RQ1})---including their retention and improvement---and fairness perceptions (\textbf{RQ2}). 
%\erase{In the first set of models, 
%we used ``{\em Fair AI}'' as the independent variable;  
%this allowed us to examine how the AI model's fairness level across groups affects all dependent variables (i.e., improvement, retention, and fairness perceptions). For the second set of models, we 
%used ``{\em Advantaged}'' and ``{\em Disadvantaged}'' as independent variables. 
%This enabled us to understand how the AI model's bias towards the specific group that the decision subject belongs to affects all dependent variables. In all of our regressions, we controlled the subject's initial credit level, sensitivity to fairness, risk attitude, and empathy level as covariates in order to improve the precision of the estimated effects of the independent variables.} 
To emphasize the exploratory nature of this study, we followed the interval estimate method~\cite{cumming2014new,dragicevic2016fair} in our analysis. That is, our regression results 
%of our regression models 
are interpreted via the
estimated coefficient values for the independent variables as well as their 95\% bootstrap confidence intervals ($R=1000$). When the 95\% bootstrap confidence interval of the coefficient for an independent variable does not include zero, we consider the effect of the independent variable to be reliable~\cite{cumming2014new}.

\subsection{Experimental Results}

368 subjects participated in our experiment and passed all filtering procedures (see Appendix~\ref{sec:demographic} for the demographics)\footnote{We conducted a priori power calculations using the G*Power software~\cite{faul2007g}. %These calculations were carried out for two distinct regression models, which are detailed in Section~\ref{sec:analysisMethods}. 
For a regression model, we assumed an effect size $f^2$ = 0.05, an $\alpha$ error probability of 0.05, and a desired power level of 0.95. Results suggest that 262 subjects are needed for achieving the desired level of treatment effect for the independent variable ``Fair AI'', while 312 subjects are needed for achieving the desired level of effect for the independent variables ``Advantaged'' and ``Disadvantaged''. Thus, in each (sub)experiment of our studies, we targeted at recruiting at least 312 subjects.}. %Among them, 54\% self-identified as male, 45\% self-identified as female, and 1\% self-identified as other genders. The most frequent racial group among our subjects was White. 
%For a complete description of the subject demographics, 
In the following, we analyzed the full dataset collected from these 368 subjects to answer our research questions.

\subsubsection{RQ1: Impacts of AI's decision fairness on engagement.}

First, we look into that in decision subjects' strategic, repeated interactions with an AI model, how the AI model's decision fairness
%, including its fairness level across groups and its favorable/unfavorable treatment to the subject's group, 
affects their engagement with the model, including their willingness to improve their qualification and their willingness to be subject to the AI model's decisions  (see Figure~\ref{fig:study1} in Appendix~\ref{sec:study1More} for the histograms of the number of improvement attempts/loan applications made by subjects in different treatments or different groups).
%, and their perceived fairness of the AI model. 

Regarding the willingness to improve their qualification, the average number of times that improvement attempts were made was $M_{\text{fair}}=4.11$ $(SD=3.43)$ and $M_{\text{unfair}}=4.11$ $(SD=3.26)$ for subjects in the fair AI and unfair AI treatments, respectively. Model 1 in Table~\ref{tab: rq1_rt_imp_fp} examines whether the AI model's fairness level {\em across groups} has any impact on subjects' willingness to improve their qualification.
Here, the estimated coefficient for the independent variable ``{\em Fair AI}'' was not reliably different from zero ($\beta=-0.03 [-0.73, 0.62]$). This suggests that subjects' average level of willingness to improve their qualification is {\em not} impacted by the AI model's decision fairness across groups.  
%as they interact with the AI model repeatedly and strategically. 
%if they can strategically respond to AI decisions in their repeated interactions with the AI model. 
Moreover, as shown in Table~\ref{tab: rq1_rt_imp_fp} (Model 2), we also find that the AI model's decision fairness {\em on the subject's own group} does not impact their willingness to improve, as neither of the coefficients associated with ``{\em Advantaged}'' and ``{\em Disadvantaged}'' are reliably different from zero.

\begin{table}[t]
\centering
%\footnotesize
\begin{adjustbox}{width=0.48\textwidth}
\begin{tabular}{c|cc||cc||cc}
\toprule
& \multicolumn{2}{c||}{\textbf{Improvement}} & \multicolumn{2}{c||}{\textbf{Retention}} & \multicolumn{2}{c}{\textbf{Perceived Fairness}} \\
\cline{2-7}
& {\em Model 1} & {\em Model 2} & {\em Model 3} & {\em Model 4} & {\em Model 5} & {\em Model 6} \\
\midrule
Fair AI & \tabincell{c}{-0.03\\\footnotesize{[-0.73, 0.62]}} & & \tabincell{c}{-0.42\\\footnotesize{[-1.18, 0.3]}} & & \tabincell{c}{0.59\\\footnotesize{[-0.47, 1.69]}} & \\
\hline
Advantaged & & \tabincell{c}{0.49\\\footnotesize{[-0.32, 1.38]}} & & \tabincell{c}{0.77\\\footnotesize{[-0.18, 1.76]}} & & \tabincell{c}{0.27\\%[0.5ex]
\footnotesize{[-0.95, 1.53]}} \\
\hline
Disadvantaged & & \tabincell{c}{-0.40\\\footnotesize{[-1.19, 0.37]}} & & \tabincell{c}{0.08\\\footnotesize{[-0.88, 1.01]}} & & \tabincell{c}{\textbf{-1.41}\textsuperscript{\S}
% \textsuperscript{\S}
\\\textbf{\footnotesize{[-2.78, -0.11]}}} \\
\hline
Risk attitude & \tabincell{c}{0.05
% \textsuperscript{***}
\\%[0.5ex]
\footnotesize [-0.03, 0.14]} & \tabincell{c}{0.05\\\footnotesize{[-0.04, 0.12]}} & \tabincell{c}{0.04\\\footnotesize{[-0.04, 0.17]}} & \tabincell{c}{0.07\\\footnotesize{[-0.05, 0.16]}} & \tabincell{c}{0.06\textsuperscript{\S}\\\footnotesize{[0.2, 0.5]}} & \tabincell{c}{0.36\textsuperscript{\S}\\\footnotesize{[0.19, 0.49]}} \\
\hline
Fairness Sensitivity & \tabincell{c}{-0.00\\\footnotesize{[-0.16, 0.14]}} & \tabincell{c}{-0.01\\\footnotesize{[-0.16, 0.14]}} & \tabincell{c}{-0.03\\\footnotesize{[-0.2, 0.14]}} & \tabincell{c}{-0.03\\\footnotesize{[-0.2, 0.13]}} & \tabincell{c}{-0.11\\\footnotesize{[-0.38, 0.12]}} & \tabincell{c}{-0.13\\%[0.5ex]
\footnotesize{[-0.38, 0.1]}} \\
\hline
Empathy & \tabincell{c}{0.13\textsuperscript{\S}\\\footnotesize{[+0.00, 0.24]}} & \tabincell{c}{0.13\textsuperscript{\S}\\\footnotesize{[+0.00, 0.24]}} & \tabincell{c}{0.12\\\footnotesize{[-0.02, 0.25]}} & \tabincell{c}{0.12\\\footnotesize{[-0.02, 0.25]}} & \tabincell{c}{-0.11\\\footnotesize{[-0.31, 0.09]}} & \tabincell{c}{-0.11\\%[0.5ex]
\footnotesize{[-0.3, 0.09]}} \\
\hline
Initial Credit Score & \tabincell{c}{-0.13\textsuperscript{\S}\\\footnotesize{[-0.26, -0.01]}} & \tabincell{c}{-0.13\textsuperscript{\S}\\\footnotesize{[-0.25, -0.00]}} & \tabincell{c}{0.22\textsuperscript{\S}\\\footnotesize{[0.07, 0.37]}} & \tabincell{c}{0.22\textsuperscript{\S}\\\footnotesize{[0.08, 0.38]}} & \tabincell{c}{0.19\textsuperscript{\S}\\\footnotesize{[+0.00, 0.37]}} & \tabincell{c}{0.20\textsuperscript{\S}\\%[0.5ex]
\footnotesize{[0.02, 0.39]}} \\
\hline
Constant & \tabincell{c}{2.49\textsuperscript{\S}\\\footnotesize{[0.13, 4.86]}} & \tabincell{c}{2.54\textsuperscript{\S}\\\footnotesize{[0.22, 4.85]}} & \tabincell{c}{3.76\textsuperscript{\S}\\\footnotesize{[1.1, 6.36]}} & \tabincell{c}{3.42\textsuperscript{\S}\\\footnotesize{[0.74, 6.07]}} & \tabincell{c}{17.51\textsuperscript{\S}\\\footnotesize{[13.34 , 21.75]}} & \tabincell{c}{18.27\textsuperscript{\S}\\%[0.5ex] 
\footnotesize{[14.19, 22.39]}} \\
\bottomrule
\end{tabular}
\end{adjustbox}
\vspace{-5pt}
\caption{Regression models predicting decision subjects' improvement, retention, and perceived fairness based on the AI model's decision fairness for Study 1. %Our results indicate that subjects who are being placed by the AI model at the disadvantaged position for receiving the favorable decision rate the model as less fair (Model 6). 
Coefficients and their 95\% bootstrap confidence intervals are reported. A superscript \textsuperscript{\S} indicates that the estimated coefficient is reliably different from zero.
%the 95\% bootstrap confidence interval does not include zero, suggesting a reliable effect. 
%Significant coefficients 
Reliable effects of the independent variables of interests 
%on variables of interest 
are bolded. %excluding the covariates and constant.
}
% \Description{

% Looking at Improvement. Empathy: In both 'Model 1' and 'Model 2', the Empathy variable shows a positive coefficient of 0.13, with a confidence interval ranging from +0.00 to +0.24, indicating a reliable effect. Initial Credit Score: In 'Model 1', the Initial Credit Score variable has a negative coefficient of -0.13, with a confidence interval from -0.26 to -0.01, suggesting a reliable effect. Looking at Retention. Initial Credit Score: In both 'Model 3' and 'Model 4', the Initial Credit Score exhibits positive coefficients of 0.22, with confidence intervals ranging from +0.07 to +0.37 and +0.08 to +0.38 respectively, showing a reliable effect. Looking at Perceived Fairness. Disadvantaged: In 'Model 6', the Disadvantaged variable displays a negative coefficient of -1.41, with a confidence interval from -2.78 to -0.11, implying that subjects who were disadvantaged by the AI model perceived it as less fair. Risk Attitude: Both 'Model 5' and 'Model 6' indicate a positive relationship with the Risk Attitude variable, with coefficients of 0.36 and 0.34, respectively, and confidence intervals ranging from +0.19 to +0.49 and +0.20 to +0.39. This suggests a reliable effect. Initial Credit Score: For 'Model 5' and 'Model 6', the Initial Credit Score has coefficients of 0.19 and 0.20, with confidence intervals ranging from +0.00 to +0.37 and +0.02 to +0.39, respectively, indicating reliable effects.

% }
\label{tab: rq1_rt_imp_fp}
\vspace{-18pt}
\end{table}

Similar observations can also be made for decision subjects' retention. On average, subjects in the fair AI treatment interacted with the AI model for $M_{\text{fair}}=6.23$ ($SD=3.93$) rounds, while subjects in the unfair AI treatment interacted with the AI model for $M_{\text{unfair}}=6.58$ ($SD=3.72$) rounds. 
%In addition, as shown in Figure~\ref{fig:study1_survival}, subjects appear to be willing to continue being subject to the AI model's decisions to a similar extent, regardless of whether the AI model places the subject's group at an advantaged or disadvantaged position. 
Regression results are shown in the middle panel of Table~\ref{tab: rq1_rt_imp_fp} (Models 3 and 4). We find that once decision subjects can strategically react to the AI model's decisions on them, 
%get the opportunities to increase their chance of receiving the favorable decision by improving their qualification, 
on average, they are equally willing to keep interacting with the AI model regardless of the model's decision fairness, both across groups and specifically towards their group. 

Interestingly, across Models 1--4 in Table~\ref{tab: rq1_rt_imp_fp}, we also notice that subjects with higher initial credit score levels consistently made fewer improvement attempts, but interacted with the AI model for more rounds. This is expected and consistent with our conjecture.

\subsubsection{RQ2: Impacts of AI's decision fairness on fairness perceptions.} 
%Next, we move on to examine the impacts of the AI model's decision fairness on decision subjects' average level of fairness perceptions of the model. 
Subjects in the fair AI treatment reported an average fairness rating of $18.77$ ($SD=5.15$) for the AI model. Meanwhile, subjects in the unfair AI treatment reported an average fairness rating of 18.02 ($SD=5.40$), with those who were placed at the advantaged position (i.e., the red group) reported an average rating of 19.0 ($SD=4.75$) and those who were placed at the disadvantaged position (i.e., the blue group) reported an average rating of 17.0 ($SD=5.81$).
%However, when  (
Table~\ref{tab: rq1_rt_imp_fp} (Models 5 and 6)
shows the results of the regression analysis. 
%, we have different findings. 
While the AI model's fairness level across different groups still does not appear to reliably influence decision subjects' average level of perceived fairness of the model (Model 5),  
%For example, while decision subjects do not seem to perceive the AI model as fairer if the AI model systematically favors the subject's group, \old{in both the sub-experiments that adopt a ``Constant'' or ``Hard to easy'' qualification improvement scheme,} 
we do find that when the AI model systematically biases against the subject's group, the subject perceives the model as less fair (i.e., $\beta=-1.41 [-2.78, -0.11]$ for ``{\em Disadvantaged}'' in Model 6). Also, consistent with our conjecture, we also observed that subjects who were more risk-seeking or had a higher initial credit score level tended to perceive the AI model as fairer.
%i.e., p-value for the variable ``disadvantaged'' in Model 6: $p=0.026$
%). 

%% file: study2.tex
%\ag{Edit Start}
In Study 1, we have answered \textbf{RQ1} and \textbf{RQ2}
%our research questions 
in an environment where the difficulty for decision subjects to improve their qualification does not vary with their current qualification levels. 
In Study 2, we
explore the generalizability of our Study 1 results in different environments where the qualification improvement difficulty varies with one's current qualification level and therefore answer \textbf{RQ3}.

\subsection{Experimental Design}

In Study 2, we conducted two sub-experiments simultaneously. The design of each sub-experiment was identical to the experiment designed for Study 1 with only one exception: In Study 1, the success rate for a subject to progress to the next credit level after they make an improvement attempt was set at a fixed value, while we vary this value in different ways in each sub-experiment of Study 2:  
%In each of the two sub-experiments of Study 2, we considered that the success rate of a credit improvement can vary with the subject's current credit level in a specific way: 
%\new{.} \new{ First, in the \emph{Easy to hard} sub-experiment, it is easier for lowly-qualified subjects to improve their qualification than highly-qualified subjects. 
%Specifically, the success rate for subjects with the lowest credit level (i.e., 300--350) in an improvement attempt was set to 80\% and the success rate decreased by 8\% for each increased level of credit score. The lowest possible success rate was 8\% for subjects with the highest credit level who could still make an improvement attempt (i.e., 750--800). Note that the average success rate across subjects of different credit levels was still 44\%, which was the same as the constant success rate 
%that in the constant qualification improvement scheme 
%used in Study 1. Second, in the \emph{Hard to easy} sub-experiment, it is more difficult for lowly-qualified subjects to improve their qualification than highly-qualified subjects. Specifically, the success rate in an improvement attempt ranged from 8\% (for subjects with the 300--350 credit level) to 80\% (for subjects with the 750--800 credit level), and it increased by 8\% for each increased level of credit score.   
%}
%\old{
\squishlist
\item \textbf{Easy to hard}: In this sub-experiment, it is easier for lowly-qualified subjects to improve their qualification than highly-qualified subjects. 
Specifically, the success rate for subjects with the lowest credit level (i.e., 300--350) in an improvement attempt was set to 80\% and the success rate decreased by 8\% for each increased level of credit score. The lowest possible success rate was 8\% for subjects with the highest credit level who could still make an improvement attempt (i.e., 750--800)\footnote{Note that the average success rate across subjects of different credit levels was still 44\%, which was the same as the constant success rate 
%that in the constant qualification improvement scheme 
used in Study 1.}. 
\item \textbf{Hard to easy}: 
In this sub-experiment, it is more difficult for lowly-qualified subjects to improve their qualification than highly-qualified subjects. Specifically, the success rate in an improvement attempt ranged from 8\% (for subjects with the 300--350 credit level) to 80\% (for subjects with the 750--800 credit level), and it increased by 8\% for each increased level of credit score.   
\squishend
%}\old{
%We used the same experimental procedure as that in Study 1 by posting the experiment as a HIT on MTurk exclusively for U.S. workers. However, there were 
The experimental procedure was the same as Study 1 except that: (1) Workers who had previously participated in the experiment in Study 1 were excluded from participating in this new experiment; (2) Workers were randomly assigned to one of the two sub-experiments as defined above.%}
%\new{Finally, the same experimental procedure as that in Study 1 was used with two key differences: (1) Workers who had previously participated in the experiment in Study 1 were excluded from participating in this new experiment; (2) Workers were randomly assigned to one of the two sub-experiments as defined above.}

\begin{table*}[t]
\centering
%\small
\begin{adjustbox}{width=1\textwidth}
\begin{tabular}{c|cc|cc||cc|cc||cc|cc}
\toprule
 & \multicolumn{4}{c||}{\textbf{Improvement}} & \multicolumn{4}{c||}{\textbf{Retention}} & \multicolumn{4}{c}{\textbf{Perceived Fairness}}\\ 
 \cline{2-13} 
  & \multicolumn{2}{c|}{\textbf{Easy to hard}} & \multicolumn{2}{c||}{\textbf{Hard to easy}} & \multicolumn{2}{c|}{\textbf{Easy to hard}} & \multicolumn{2}{c||}{\textbf{Hard to easy}} & \multicolumn{2}{c|}{\textbf{Easy to hard}} & \multicolumn{2}{c}{\textbf{Hard to easy}}\\
 \cline{2-13}
 & {\em Model 1} & {\em Model 2} & {\em Model 3} & {\em Model 4}  &  {\em Model 5} & {\em Model 6} & {\em Model 7} & {\em Model 8} & {\em Model 9} & {\em Model 10} & {\em Model 11} & {\em Model 12}\\
\midrule
Fair AI & \tabincell{c}{0.18\\\footnotesize{[-0.56, 0.96]}} & & \tabincell{c}{-0.00\\\footnotesize{[-0.57, 0.57]}} & & \tabincell{c}{0.29\\\footnotesize{[-0.47, 1.14]}} & & \tabincell{c}{0.01\\\footnotesize{[-0.68, 0.7]}} & & \tabincell{c}{0.34\\\footnotesize{[-0.58, 1.31]}} & & \tabincell{c}{\textbf{1.45}\textsuperscript{\S}\\\textbf{\footnotesize{[0.5, 2.46]}}} & \\
\hline
Advantaged & & \tabincell{c}{0.11\\\footnotesize{[-0.9, 1]}} & & \tabincell{c}{-0.30\\\footnotesize{[-0.99, 0.42]}} & & \tabincell{c}{0.18\\\footnotesize{[-0.84, 1.1]}} & & \tabincell{c}{-0.18\\\footnotesize{[-1.08, 0.66]}} & & \tabincell{c}{-0.06\\\footnotesize{[-1.3, 1.17]}} & & \tabincell{c}{-0.58\\\footnotesize{[-1.89, 0.64]}} \\
\hline
Disadvantaged & & \tabincell{c}{-0.46\\\footnotesize{[-1.44, 0.45]}} & & \tabincell{c}{0.24\\\footnotesize{[-0.47, 0.94]}} & & \tabincell{c}{-0.79\\\footnotesize{[-1.78, 0.19]}} & & \tabincell{c}{0.12\\\footnotesize{[-0.7 , 0.97]}} & & \tabincell{c}{-0.63\\\footnotesize{[-1.94, 0.61]}} & & \tabincell{c}{\textbf{-2.14}\textsuperscript{\S}\\\textbf{\footnotesize{[-3.35, -0.89]}}} \\ 
\hline
Risk attitude & \tabincell{c}{0.13\textsuperscript{\S}\\ \footnotesize{[0.04, 0.24]}} & \tabincell{c}{0.14\textsuperscript{\S}\\\footnotesize{[0.05 , 0.24]}} & \tabincell{c}{0.08\textsuperscript{\S}\\\footnotesize{[+0.00, 0.17]}} & \tabincell{c}{0.08\\\footnotesize{[-0.00, 0.17]}} & \tabincell{c}{0.15\textsuperscript{\S}\\\footnotesize{[0.05, 0.25]}} & \tabincell{c}{0.16\textsuperscript{\S}\\\footnotesize{[0.06, 0.26]}} & \tabincell{c}{0.18\textsuperscript{\S}\\\footnotesize{[0.09, 0.28]}} & \tabincell{c}{0.18\textsuperscript{\S}\\\footnotesize{[0.08, 0.28]}} & \tabincell{c}{0.34\textsuperscript{\S}\\\footnotesize{[0.2, 0.49]}} & \tabincell{c}{0.35\textsuperscript{\S}\\\footnotesize{[0.21, 0.49]}} & \tabincell{c}{0.31\textsuperscript{\S}\\\footnotesize{[0.19, 0.44]}} & \tabincell{c}{0.32\textsuperscript{\S}\\\footnotesize{[0.19, 0.44]}}\\
\hline
Fairness Sensitivity & \tabincell{c}{-0.10\\\footnotesize{[-0.25, 0.03]}} & \tabincell{c}{-0.10\\\footnotesize{[-0.26, 0.04]}} & \tabincell{c}{-0.03\\\footnotesize{[-0.16, 0.09]}} & \tabincell{c}{-0.03\\\footnotesize{[-0.16, 0.09]}} & \tabincell{c}{-0.13\textsuperscript{\S}\\\footnotesize{[-0.28 , -0.00]}} & \tabincell{c}{-0.14\textsuperscript{\S}\\\footnotesize{[-0.28, -0.01]}} & \tabincell{c}{-0.04\\\footnotesize{[-0.19, 0.11]}} & \tabincell{c}{-0.04\\\footnotesize{[-0.19, 0.11]}}  & \tabincell{c}{0.25\textsuperscript{\S}\\\footnotesize{[0.05, 0.51]}} & \tabincell{c}{0.25\textsuperscript{\S}\\\footnotesize{[0.04, 0.51]}} & \tabincell{c}{0.10\\\footnotesize{[-0.11, 0.31]}} & \tabincell{c}{0.11\\\footnotesize{[-0.11, 0.31]}}\\
\hline
Empathy & \tabincell{c}{0.13\textsuperscript{\S}\\\footnotesize{[+0.00, 0.25]}} & \tabincell{c}{0.12\\\footnotesize{[ -0.00, 0.24]}} & \tabincell{c}{0.10\\\footnotesize{[-0.00, 0.22]}} & \tabincell{c}{0.10\\\footnotesize{[-0.00, 0.22]}} & \tabincell{c}{0.16\textsuperscript{\S}\\\footnotesize{[0.03 , 0.29]}} & \tabincell{c}{0.14\textsuperscript{\S}\\\footnotesize{[0.02, 0.28]}} & \tabincell{c}{0.12\textsuperscript{\S}\\\footnotesize{[+0.00 , 0.27]}} & \tabincell{c}{0.12\textsuperscript{\S}\\\footnotesize{[+0.00, 0.27]}} & \tabincell{c}{-0.14\\\footnotesize{[-0.34, 0.06]}} & \tabincell{c}{-0.15\\\footnotesize{[-0.35, 0.05]}} & \tabincell{c}{-0.20\textsuperscript{\S}\\\footnotesize{[-0.35, -0.03]}} & \tabincell{c}{-0.20\textsuperscript{\S}\\\footnotesize{[-0.36, -0.03]}} \\
\hline
Initial Credit Score & \tabincell{c}{-0.11\\\footnotesize{[-0.26, 0.04]}} & \tabincell{c}{-0.11\\\footnotesize{[-0.27, 0.03]}} & \tabincell{c}{-0.24\textsuperscript{\S}\\\footnotesize{[-0.34, -0.13]}} & \tabincell{c}{-0.24\textsuperscript{\S}\\\footnotesize{[-0.34, -0.12]}} & \tabincell{c}{0.16\textsuperscript{\S}\\\footnotesize{[0.01, 0.32]}} & \tabincell{c}{0.16\textsuperscript{\S}\\\footnotesize{[0.01, 0.31]}} & \tabincell{c}{0.25\textsuperscript{\S}\\\footnotesize{[0.12, 0.38]}} & \tabincell{c}{0.25\textsuperscript{\S}\\\footnotesize{[0.12, 0.39]}} & \tabincell{c}{0.13\\\footnotesize{[-0.08, 0.34]}} & \tabincell{c}{0.13\\\footnotesize{[-0.08, 0.33]}} & \tabincell{c}{0.24\textsuperscript{\S}\\\footnotesize{[0.03, 0.41]}} & \tabincell{c}{0.23\textsuperscript{\S}\\\footnotesize{[0.03, 0.4]}} \\
\hline
Constant & \tabincell{c}{3.48\textsuperscript{\S}\\\footnotesize{[0.98, 5.89]}} & \tabincell{c}{3.79\textsuperscript{\S}\\\footnotesize{[1.3, 6.23]}} & \tabincell{c}{2.95\textsuperscript{\S}\\\footnotesize{[0.97, 4.91]}} & \tabincell{c}{2.94\textsuperscript{\S}\\\footnotesize{[0.99, 4.84]}} & \tabincell{c}{4.19\textsuperscript{\S}\\\footnotesize{[1.56, 6.74]}} & \tabincell{c}{4.71\textsuperscript{\S}\\\footnotesize{[2.1, 7.15]}} & \tabincell{c}{2.78\textsuperscript{\S}\\\footnotesize{[0.21, 5.06]}} & \tabincell{c}{2.78\textsuperscript{\S}\\\footnotesize{[0.31, 4.99]}} & \tabincell{c}{14.64\textsuperscript{\S}\\\footnotesize{[11.23, 18.23]}} & \tabincell{c}{15.12\textsuperscript{\S}\\\footnotesize{[11.64, 16.81]}} & \tabincell{c}{16.23\textsuperscript{\S}\\\footnotesize{[13.01, 19.36]}} & \tabincell{c}{17.71\textsuperscript{\S}\\\footnotesize{[14.41, 20.71]}}\\
\bottomrule
\end{tabular}
\end{adjustbox}
\vspace{-5pt}
\caption{Regression models predicting decision subjects' improvement, retention, and perceived fairness based on the AI model's decision fairness
in the two sub-experiments of Study 2. 
%in the respective sub-experiments for each of the two qualification improvement schemes. 
%Consistent with Study 1, our findings also indicate here that subjects disadvantaged by the AI model appear to rate the model as less fair (Model 12). 
%Our results indicate that subjects' fairness perceptions of the AI model are reliably affected by its fairness properties when it is more difficult for subjects with low qualification to improve their chance of the favorable decision (Models 11--12).
Coefficients and their 95\% bootstrap confidence intervals are reported. A superscript \textsuperscript{\S} indicates that the estimated coefficient is reliably different from zero.
%the 95\% bootstrap confidence interval does not include zero, suggesting a reliable effect. 
%Significant coefficients 
Reliable effects of the independent variables of interests 
%on variables of interest 
are bolded. %excluding the covariates and constant.
%\ag{DONE!}
%, excluding the covariates and constant.
}
\label{tab: rq3_rt_imp_fp}
\vspace{-18pt}
\end{table*}

\subsection{Experimental Results}  

A total of 713 
subjects participated in our experiment and passed all filtering procedures (``Easy to hard'': 328 subjects, ``Hard to easy'': 385 subjects). Demographic details of these subjects are reported in Appendix~\ref{sec:demographic}. %the %\hyperref[sec:appendix]{appendix}. 
Below, we analyzed the full dataset collected from these subjects to %examine the generalizability of the results of Study 1, 
answer \textbf{RQ3} by re-examining \textbf{RQ1}--\textbf{RQ2} on this dataset 
using the same analysis methods as in Study 1. 

\subsubsection{Re-examining RQ1: Impacts of AI's decision fairness on engagement.}
%We repeat the analysis on how the AI model's decision fairness affects decision subjects' willingness to improve and willingness to be subject to the model's decisions %for the two sub-experiments adopting ``Easy to hard'' and ``Hard to easy'' qualification improvement schemes 
%within each sub-experiment 
%separately (see the \hyperref[sec:appendix]{appendix} for figures of the improvement and retention distributions for subjects in the two sub-experiments).
For the ``Easy-to-hard'' sub-experiment, on average, subjects made 4.86 ($SD=3.63$) improvement attempts and stayed for 7 ($SD=3.77$) rounds in the loan application tasks when they interacted with the fair AI model. Meanwhile, subjects in the unfair AI model treatment made an average of 4.70 ($SD=3.49$) improvement attempts and interacted with the AI model for 6.67 ($SD=3.68$) rounds. 
%(see Appendix~\ref{sec:easy2Hard} for figures of subjects' improvement and retention distributions).
The regression results, as reported in Table~\ref{tab: rq3_rt_imp_fp} (Models 1, 2, 5, 6), suggest that in this sub-experiment, decision subjects' engagement with the AI model was not reliably influenced by either the AI model's decision fairness across groups or on the specific group that the subject belonged to. This is also true for subjects in the ``Hard-to-easy'' sub-experiment---%Here, an average subject in the fair AI model treatment attempted to improve their qualification in 3.68 ($SD=3.14$) rounds and stayed in the tasks for 6.44 ($SD=3.84$) rounds, which was not substantially different from an average subject in the unfair AI model treatment (improvement: $M=3.76, SD=2.97$; retention: $M=6.43, SD=3.67$). 
Again, in the regression models learned for this sub-experiment (Models 3, 4, 7, 8 in Table~\ref{tab: rq3_rt_imp_fp}), we did not detect any reliable effect of the AI model's decision fairness on subjects' willingness to improve their qualification or continue interacting with the model (see Appendix~\ref{sec:easy2Hard} and~\ref{sec:hard2Easy} for additional figures and statistics).

%Consistent with our findings in Study 1, it is found that, on average, decision subjects' willingness to improve themselves and willingness to keep interacting with the AI model
%are not significantly influenced by the AI model's level of fairness across groups or its decisions regarding the subject's group, regardless of how the difficulty of qualification improvement changes with the subject's current qualification level. 

\subsubsection{Re-examining RQ2: Impacts of AI's decision fairness on fairness perceptions.} %However, when examining the impact of the AI model's decision fairness on decision subjects' fairness perceptions (
Table~\ref{tab: rq3_rt_imp_fp} (Models 9--12) shows how the AI model's decision fairness affects %the average level of 
decision subjects' perceived fairness of the AI model in the two sub-experiments of Study 2. 
%Interestingly, we find that %when it is easier for lowly-qualified subjects to improve their qualification level than highly-qualified subjects (i.e., 
We note that in the ``Easy to hard'' sub-experiment, 
subjects' perceived fairness of the AI model is {\em not} reliably impacted by either the AI model's group-level fairness or the AI model's bias on the subject's own group.  
%(e.g., %average perceived fairness--
%fair AI treatment: $M=19.05$, $SD=4.71$; unfair AI treatment, red group: $M=18.59$, $SD=5.25$; unfair AI treatment, blue group: $M=18.92$, $SD=4.99$). 
In contrast, in the ``Hard to easy'' sub-experiment, 
we find that subjects' perceived fairness of the AI model is affected by the AI model's fairness properties (e.g., %average perceived fairness of AI: 
fair AI treatment: $M=19.37$, $SD=4.46$; unfair AI treatment, red group: $M=18.75$, $SD=5.34$; unfair AI treatment, blue group: $M=17.26$, $SD=5.75$). In particular, in this sub-experiment, decision subjects' average perceived fairness of the AI model increased when the AI model made fair decisions across different groups (i.e., the coefficient for ``{\em Fair AI}'' in Model 11 is $\beta=1.45 [0.5, 2.46]$). In addition, those subjects who have been placed at the disadvantaged position by the AI model also perceived the model to be %significantly 
less fair (i.e., 
the coefficient for ``{\em Disadvantaged}'' in Model 12 is $\beta=-2.14 [-3.35, -0.89]$).

%% file: study3.tex
Finally, to answer \textbf{RQ4}, we conduct  Study 3 where the AI model's fairness %properties 
is examined with respect to a protected attribute, %i.e., 
gender, and subjects' group identities in the study 
were determined by their self-reported, real-world gender. 

\subsection{Experimental Design}

In Study 3, we adopted the experimental design from Study 1 and made a few minor changes. 
First, instead of assigning a fictional group identity (red or blue) to each subject, we asked subjects to self-report their gender at the beginning of the experiment and used it as their group identity in the experiment. 
Second, in the ``fair AI'' treatment, the AI model employed provided equal decisions to male and female subjects in granting loans; the approval rate of this AI model to both male and female was determined by Table~\ref{tab:fair}. In contrast, in the ``unfair AI'' treatment, the AI model exhibited gender bias and favored male over female subjects in granting loans, and its approval rate for male and female was determined by Table~\ref{tab:unfair:red} and Table~\ref{tab:unfair:blue}, respectively.  
%(i.e., males correspond to the red group in Study 1, and females correspond to the blue group in Study 1)\footnote{\new{This unfair AI model was designed to mirror the AI model's gender biases against females in the real world that were increasingly revealed by researchers~\cite{dastin_2018,bolukbasi2016man,buolamwini2018gender,nadeem2020gender}.}}.  
This unfair AI model was designed to mirror the AI model's gender biases against females in the real world that were increasingly revealed by researchers~\cite{dastin_2018,bolukbasi2016man,buolamwini2018gender,nadeem2020gender}. 
Finally, 
we also modified the flowcharts that were shown to subjects at the end of each round, so that they  summarized the AI model's decisions on simulated loan applicants who were grouped by their gender (instead of the red vs. blue group identities as used in Study 1).

The procedure of Study 3 was also largely the same as Study 1, except for a few minor changes: 
(1) Previous participants from Studies 1 and 2 were excluded. (2) The AI model, as discussed earlier, determined loan approvals based on the subject's self-reported gender and their credit level. (3) We also incorporated a deceptive component in the study to caputre subjects' genuine reactions towards a possibly biased AI model while avoiding actually paying female subjects systematically less in our experiment.  
In particular, at the beginning of the experiment, subjects were told that %beyond the \$2 base payment, 
their bonus payment %that they would receive from 
in the experiment was proportional to the final balance in their account, 
with every 500 coins translating to \(\$1.50\). However, in reality, to prevent gender bias in study payments, each subject was given 
a fixed bonus of \(\$3.30\). 
In other words, all subjects of Study 3 eventually received a total payment of \(\$5.30\). 
%\footnote{\new{This deception was used to make subjects believe that their final earnings were directly influenced by the AI model's decisions based on their real-world gender, thus enabling us to capture subjects' genuine reactions towards an AI model that may or may not exhibit gender bias. Upon completing the experiment, subjects were immediately debriefed about the deception they had experienced. 
%The median time subjects spent on the experiment was 19 minutes, which resulted in an hourly rate of approximately \$16.70. %\old{This deception was used to make subjects believe that their final earnings were directly influenced by the AI model's decisions based on their real-world gender, thus enabling us to capture subjects' genuine reactions towards an AI model that may or may not exhibit gender bias. Upon completing the experiment, subjects were immediately debriefed about the deception they had experienced. 
%The median time subjects spent on the experiment was 19 minutes, which resulted in an hourly rate of approximately \$16.70.}

\subsection{Experimental Results}

A total of 416 subjects participated in Study 3 and passed the filtering procedure. Among them, 55\% self-identified as male, and 45\% self-identified as female (see Appendix~\ref{sec:demographic} for more demographics). %the \hyperref[sec:appendix]{appendix} for the breakdowns on other demographics). 
In the following, we used the full dataset from these subjects to re-do the same kind of analysis that we conducted for \textbf{RQ1--RQ2}, in order to examine whether our results of Study 1 still hold true when the AI model's fairness properties was examined with respect to a protected social attribute, i.e., gender (\textbf{RQ4}).
%gender and subjects' group identity in the experiment was decided by their real-world gender. 

\subsubsection{Re-examining RQ1: Impacts of AI's decision fairness on engagement.}
%We first revisited the analysis to determine the impact of the AI model's fairness on decision subjects' willingness to improve and willingness to be subject to the model's decisions, when the AI model's fairness properties are discussed with respect to decision subjects' real-world gender. 
%now accounting for their real-world genders. 
Regarding subjects' engagement with the AI model in this study, we found that male subjects made 4.28 ($SD=3.44$) improvement attempts and stayed for 6.83 ($SD=3.73$) rounds on average, while female subjects made 3.98 ($SD=3.17$) improvement attempts and stayed for 6.46 ($SD=3.82$) rounds.
The regression results are presented in Table~\ref{tab: gender_rq1} (Models 1--4). 
%As in line with our results from Studies 1 and 2, 
Again, we find that, on average, decision subjects' willingness to improve themselves and willingness to keep interacting with the AI model are not reliably affected by the AI model's fairness level across groups or its decision fairness towards the subjects' group. In other words, our Study 1 results still hold true even when the AI model's fairness is examined with respect to salient protected attributes like gender. 
%with the inclusion of 
%real-world gender.

\subsubsection{Re-examining RQ2: Impacts of AI's decision fairness on fairness perceptions.}
%We then investigated into the influence of the AI model's decision fairness on the average level of decision subjects' perceived fairness of the AI model. 
In Study 3, the average rating of the perceived fairness of the AI model was 18.90 ($SD=5.33$) for subjects in the fair AI model treatment, and 18.24 ($SD=5.75$) for subjects in the unfair AI model treatment. Within the unfair AI model treatment, the average perceived fairness rating of the AI model was 18.92 ($SD=5.64$) and 17.30 ($SD=5.79$) for the male and female subjects, respectively.
The regression results, as presented in Table~\ref{tab: gender_rq1} (Models 5 and 6), 
%. Consistent with our findings from Study 1, we 
again suggest that female subjects 
%who were being placed in a disadvantaged position by the AI model 
perceived the AI model as less fair if the AI model systematically biased against them. This is evidenced by the reliably negative coefficient estimated for the independent variable ``{\em Disadvantaged}'' in Model 6 (i.e., 
%p-value for the variable ``disadvantaged'' in Model 6: $p=0.034$).
$\beta=-1.38 [-2.65, -0.16]$).
 In contrast, male subjects who were being placed at an advantaged position by the unfair AI model did not show any decrease in their perceived fairness of the model, despite the AI model exhibited a clear gender bias. 

\begin{table}[t]
\centering
\begin{adjustbox}{width=0.48\textwidth}
\begin{tabular}{c|cc||cc||cc}
\toprule
& \multicolumn{2}{c||}{\textbf{Improvement}} & \multicolumn{2}{c||}{\textbf{Retention}} & \multicolumn{2}{c}{\textbf{Perceived Fairness}} \\
\cline{2-7}
& {\em Model 1} & {\em Model 2} & {\em Model 3} & {\em Model 4} & {\em Model 5} & {\em Model 6} \\
\midrule
Fair AI & \tabincell{c}{0.12\\\footnotesize{[-0.49, 0.78]}} & & \tabincell{c}{-0.45\\\footnotesize{[-1.17, 0.35]}} & & \tabincell{c}{0.45\\\footnotesize{[-0.52, 1.46]}} & \\
\hline
Advantaged & & \tabincell{c}{-0.04\\\footnotesize{[-0.83, 0.68]}} & & \tabincell{c}{0.64\\\footnotesize{[-0.24, 1.46]}} & & \tabincell{c}{0.23\\\footnotesize{[-1, 1.41]}} \\
\hline
Disadvantaged & & \tabincell{c}{-0.22\\\footnotesize{[-1.08, 0.5]}} & & \tabincell{c}{0.18\\\footnotesize{[-0.8, 1.08]}} & & \tabincell{c}{\textbf{-1.38}\textsuperscript{\S}\\\textbf{\footnotesize{[-2.65, -0.16]}}} \\
\hline
Risk attitude & \tabincell{c}{0.09\textsuperscript{\S}\\\footnotesize{[0.01, 0.17]}} & \tabincell{c}{0.09\textsuperscript{\S}\\\footnotesize{[0.01, 0.17]}} & \tabincell{c}{0.13\textsuperscript{\S}\\\footnotesize{[0.04, 0.22]}} & \tabincell{c}{0.13\textsuperscript{\S}\\\footnotesize{[0.04, 0.22]}} & \tabincell{c}{0.41\textsuperscript{\S}\\\footnotesize{[0.28, 0.54]}} & \tabincell{c}{0.41\textsuperscript{\S}\\\footnotesize{[0.27, 0.54]}} \\
\hline
Fairness Sensitivity & \tabincell{c}{0.04\\\footnotesize{[-0.1, 0.17]}} & \tabincell{c}{0.04\\\footnotesize{[-0.1 , 0.17]}} & \tabincell{c}{0.02\\\footnotesize{[-0.14, 0.17]}} & \tabincell{c}{0.02\\\footnotesize{[-0.14, 0.17]}} & \tabincell{c}{-0.14\\\footnotesize{[-0.38, 0.12]}} & \tabincell{c}{-0.13\\\footnotesize{[-0.37, 0.13]}} \\
\hline
Empathy & \tabincell{c}{0.05\\\footnotesize{[-0.05, 0.16]}} & \tabincell{c}{0.05\\\footnotesize{[-0.05, 0.16]}} & \tabincell{c}{0.01\\\footnotesize{[-0.11 , 0.13]}} & \tabincell{c}{0.01\\\footnotesize{[-0.11, 0.13]}} & \tabincell{c}{-0.01\\\footnotesize{[-0.19, 0.16]}} & \tabincell{c}{-0.01\\\footnotesize{[-0.18, 0.16]}} \\
\hline
Initial Credit Score & \tabincell{c}{-0.23\textsuperscript{\S}\\\footnotesize{[-0.35, -0.11]}} & \tabincell{c}{-0.23\textsuperscript{\S}\\\footnotesize{[-0.35, -0.11]}} & \tabincell{c}{0.11\\\footnotesize{[-0.03, 0.26]}} & \tabincell{c}{0.12\\\footnotesize{[-0.03, 0.26]}} & \tabincell{c}{0.36\textsuperscript{\S}\\\footnotesize{[0.17, 0.56]}} & \tabincell{c}{0.37\textsuperscript{\S}\\\footnotesize{[0.18, 0.56]}} \\
\hline
Constant & \tabincell{c}{3.14\textsuperscript{\S}\\\footnotesize{[0.9, 5.22]}} & \tabincell{c}{3.23\textsuperscript{\S}\\\footnotesize{[1.06, 5.33]}} & \tabincell{c}{5.04\textsuperscript{\S}\\\footnotesize{[2.62, 7.55]}} & \tabincell{c}{4.53\textsuperscript{\S}\\\footnotesize{[2.17, 7.06]}} & \tabincell{c}{15.21\textsuperscript{\S}\\\footnotesize{[11.43, 19.09]}} & \tabincell{c}{15.44\textsuperscript{\S}\\\footnotesize{[11.73, 19.28]}} \\
\bottomrule
\end{tabular}
\end{adjustbox}
\vspace{-5pt}
\caption{Regression models predicting decision subjects' improvement, retention, and perceived fairness based on the AI model's decision fairness for Study 3. 
%Consistent with earlier studies, our findings also indicate that subjects disadvantaged by the AI model rate the model as less fair (Model 6). 
Coefficients and their 95\% bootstrap confidence intervals are reported. A superscript \textsuperscript{\S} indicates that the estimated coefficient is reliably different from zero.
%the 95\% bootstrap confidence interval does not include zero, suggesting a reliable effect. 
%Significant coefficients 
Reliable effects of the independent variables of interests 
%on variables of interest 
are bolded. %excluding the covariates and constant.
%\ag{DONE!}
%, excluding the covariates and constant.
}
% \Description{
% Results Overview: Examining Improvement. Risk Attitude: In both 'Model 1' and 'Model 2', Risk Attitude demonstrates a positive relationship with coefficients of 0.09, accompanied by confidence intervals of [0.01, 0.17], indicating a reliable effect. Initial Credit Score: This variable shows a negative relationship with improvement in both 'Model 1' and 'Model 2', with coefficients of -0.23 and confidence intervals of [-0.35, -0.11], suggesting a reliable effect. Examining Retention. Risk Attitude: In 'Model 3' and 'Model 4', Risk Attitude maintains a positive relationship with retention, evidenced by coefficients of 0.13 and confidence intervals of [0.04, 0.22], indicating a reliable effect. Examining Perceived Fairness. Disadvantaged: 'Model 6' reveals a strong negative relationship, with a coefficient of -1.38 and a confidence interval of [-2.65, -0.16], suggesting that subjects disadvantaged by the AI model perceive it as less fair. Risk Attitude: In both 'Model 5' and 'Model 6', Risk Attitude shows positive coefficients of 0.41, with confidence intervals of [0.28, 0.54] and [0.27, 0.54] respectively, indicating a reliable effect. Initial Credit Score: 'Model 5' and 'Model 6' demonstrate strong positive relationships with coefficients of 0.36 and 0.37, and confidence intervals of [0.17, 0.56] and [0.18, 0.56] respectively, again suggesting reliable effects.

% }
\label{tab: gender_rq1}
\vspace{-20pt}
\end{table}

%% file: discussion.tex
In this paper, we conducted exploratory randomized human-subject experiments to investigate how an AI model's decision fairness affect decision subjects' engagement with and fairness perceptions of the  model when they could repeatedly and strategically respond to the AI model's decisions on them.
%\erase{Our findings suggest that in general, the AI model's fairness properties, in terms of both whether the AI model provides equal decisions to different groups in a similar way and whether the AI model favors/disfavors the specific group that the subject belongs to, have limited impact on decision subjects' willingness to improve their qualification or keep interacting with the AI model. This is true even when the AI model's fairness properties are examined with respect to salient protected attributes like gender. 
%However, the AI model's fairness properties, especially its biased decisions against a decision subject's own group, still tend to result in a decrease in the subject's perceived fairness of the model. The impact of the decision fairness of AI on decision subjects' perceived fairness of AI also appears to be larger in an environment where improving one's chance of receiving the favorable decision is particularly difficult for those who have low qualification to begin with. 
A key finding of our study is %suggest 
that when decision subjects can repeatedly and strategically respond to AI decisions, their engagement with the AI model does {\em not} reflect either the AI model's decision fairness---as decided by various fairness definitions---or their perceived fairness of the AI model. %}
In this section, we reflect on our findings, and address the limitations and future work of our study.

\vspace{2pt}
\noindent \textbf{Similarity and difference of our results with earlier findings. }
%\ag{Previous research involving decision subjects' interactions with AI systems~\cite{haiyiPaper} suggests that perceptions of fairness in AI systems are higher if the outcomes are favorable to the users or if the AI models are developed and evaluated with transparency, taking into account the representativeness of training data and the availability of information about the AI's functionality. These perceptions are influenced by individual differences such as age, gender, educational background, and cultural context; for instance, individuals with more education may have a deeper understanding of the intricacies of algorithmic decisions. In another study~\cite{altugPaper}, decision subjects who belong to groups favored by the AI system or those with a willingness for risk-taking are more likely to regard the system as fair, while those with a high sensitivity to fairness are more likely to rate the AI system as less fair. The paper~\cite{yurrita2023disentangling} underscores the significance of explanations and contestability in shaping fairness perceptions in AI, showing that explanations enhance informational fairness, while contestability improves procedural fairness. Interestingly, human oversight does not significantly affect fairness perceptions. The study also finds that these factors contribute independently to overall fairness, regardless of the decision's stakes, and highlights challenges in balancing informative clarity and efficient, human-inclusive decision-making in AI systems.} 
In this study, we find that when decision subjects can repeatedly and strategically respond to AI decisions, 
their perceived fairness of the model is 
still affected by 
the model's biases 
against their {\em own} group. %That is, decision subjects \new{they} tend to perceive an AI model as less fair if it consistently places their group at a disadvantaged position for the favorable decision. 
This is largely consistent with previous findings~\cite{haiyiPaper,altugPaper}, which suggests that decision subjects have a degree of ``outcome favorability bias'' %(i.e., AI is unfair if it grants few favorable decisions to one's own group) 
in their fairness perceptions of AI models~\cite{haiyiPaper}. This also indicates that social comparison across groups is still a key factor that contributes to decision subjects' perceived fairness of an AI model, despite the possibility to improve one's qualification  
may have shifted some of their attention towards temporal self-comparison. Indeed, in subjects' open-ended text responses in the exit survey explaining why they perceived the AI model as fair or unfair, we find evidence suggesting decision subjects' perceived fairness of the AI model may be affected by both social comparison (e.g., ``{\em people in red and blue groups with the exact same credit score ranges had different chances of getting approved}'') and temporal self-comparison (``{\em when my credit score improved, I kept getting approved for loans}''). See Appendix~\ref{sec:fairnessFactors} for more detailed analysis. 
%detailed analysis  
%factors that we identified from subjects' open-text responses that influence their perceived fairness of the AI model.
%\erase{Indeed, if temporal self-comparison was the primary driving factor of decision subjects' perceived fairness of AI models in our experiments, we would have seen subjects in all treatments and all groups perceived the AI model to be equally fair (because the improvement of qualification always results in the same amount of increase in AI's loan approval rate regardless of the subject's treatment/group assignment). In this sense, our results align with previous findings suggesting that people's sense of fairness is highly correlated with their tendency to compare themselves with others~\cite{zhen2016tend}.} 

On the other hand,
in terms of decision subjects' engagement with the AI model, our study shows that their willingness to keep interacting with the model is not affected by the AI model's biases towards/against their own group. This is different from findings in prior work~\cite{altugPaper} when the only strategic action decision subjects could take is to stop interacting with the AI model.
We speculate that this is because in our study,  the qualification improvement opportunities %provide decision subjects the possibility to utilize these opportunities to ``gain something'' (e.g., earn more favorable decisions) rather than leaving the system with nothing. In other words, it 
may have 
led decision subjects to prioritize ``utility-related considerations'' over ``fairness-related considerations'' when deciding how to engage with the AI model. 
Indeed, for those decision subjects whose group is disfavored %placed at the disadvantaged position 
by the AI model, continuing to improve themselves and interact with the AI model  
but receiving less favorable decisions from AI is certainly not ideal---as reflected in their perceived fairness of the AI model---but from a pure utility point of view, it may be better 
than simply ``boycotting'' the AI model and leaving the system with nothing.  
\vspace{2pt}
\noindent \textbf{Explaining findings of Study 2 and 3}. 
In Study 2, %\erase{as we change how the difficulty for one to improve their qualification varies with their current qualification level, we generally find the AI's decision fairness still does not have significant impacts on decision subjects' engagement with the model. However,} we note some deviations in decision subjects' fairness perceptions of the model. 
we find that the AI model's decision fairness significantly affected subjects' perceived fairness of AI only %The AI model's \erase{unequal decisions on different groups or its} biased decisions against a decision subject's own group significantly decreases the perceived fairness of the AI model 
in the ``Hard to easy'' sub-experiment, but not in the ``Easy to hard'' sub-experiment. %\erase{Here, we provide some conjectures on why we have these observations.} 
We speculate that this is because subjects' perception of an AI model's fairness is influenced by both the model's decisions on their present-self, and the model's anticipated decisions on their ideal future-self after they improve their qualification. For subjects with high initial qualifications, their primary focus might be on their present-self (as they have limited room for further improvement), hence their perceived fairness of AI might not be significantly affected by the difficulty of qualification improvement. %\erase{This focus on the present-self may imply that for these highly-qualified subjects, their perceived fairness of the AI model would not be significantly affected by how easy or how difficult it is for them to further improve their qualification.
%However, this may not be true for subjects starting with low qualification.} 
However, for subjects with low initial qualifications, 
the emphasis they put on their present-self and future-self might differ based on their belief in whether they could successfully achieve their ideal future-self.
For example, in the ``Easy to hard'' sub-experiment, subjects with lower qualifications %experienced relatively high success rate in improving their qualification and 
might be more optimistic about reaching their ideal future-self, hence they may focus more on temporal self-comparison when assessing the AI model's fairness. %\erase{This may shift subjects' focus more to temporal self-comparison when assessing the AI model's fairness, and may make subjects feel good about themselves in general---both may lead to these subjects' lower sensitivity to the biases in the AI model's decisions.}
In contrast, in the ``Hard to easy'' sub-experiment, subjects with lower qualifications found it difficult to improve their qualification. Thus, they might focus more on their present-self, engaging in social comparisons, and have a heightened sensitivity for the AI model's biases. 

%\old{and they} might generally feel less positive about themselves \new{and have a heightened sensitivity for the AI model's biases}. \erase{As such, lowly-qualified subjects in the ``Hard to easy'' sub-experiment might have heightened sensitivity to the AI model's biases, resulting both the AI model's fairness across groups and its bias against the subject's own group to substantially affect their perceived fairness of the AI model.}

%\subsubsection{Understanding the findings when AI's decision fairness is examined with respect to protected attributes}

Moreover, in Study 3, we found that decision subjects' engagement with AI was not affected by the AI model's decision fairness even when fairness is examined on subjects' real-world gender groups. %even when we incorporated subjects' real-world gender as their group identities in the experiment and examined the AI model's fairness properties with respect to groups defined by gender, our findings were still largely consistent with those of Studies 1 and 2. 
%\erase{In particular, while female subjects considered the AI model as less fair when being placed at the disadvantaged position in receiving the favorable decision by the model, they showed similar levels of willingness to engage with the model as male subjects. We provide a few possible explanations for this observation. } 
%\erase{First,} 
Prior research has pointed out the frequent challenges women encounter due to societal biases in the real world~\cite{dastin_2018,moss2012science,elsesser2016gender,lee2015gender}. Thus, we speculate that female subjects in our study might find the gender biases exhibited by the AI model in our experiment to be ``familiar''. %Despite being disappointed by it, 
Thus, they may have decided to adapt to such bias based on their past experiences of dealing with real-world bias in order to maximize the utility they may obtain from the model, especially given that they see little possibility of changing how the model works by disengaging with the unfair AI model.
Meanwhile, we also found that %when male subjects were placed at the advantaged position by an AI model that exhibited a clear gender bias, 
male subjects in Study 3 did not consider the AI model as less fair when they were placed at the advantaged position by an AI model with gender bias. 
%did not report a lower level of perceived fairness of the model compared to male subjects who encountered an AI model with no gender bias. 
This again suggests that male subjects may have suffered from the outcome favorability bias when forming their fairness perceptions of an AI model~\cite{haiyiPaper}, 
%As an implication, this finding 
and highlights the importance of identifying a representative and diverse population of decision subjects when evaluating the perceived fairness of AI models. 

\vspace{2pt}
\noindent \textbf{Implications of our findings.} 
%Findings of our study hold important implications for the real-world deployment of AI-based decision systems. First,
%Our results reveal that when decision subjects could strategically and repeatedly respond to the decisions on them made by an AI model, their overall engagement with the AI model is similar no matter whether the AI model is biased towards or against their own group. This is observed despite that subjects generally consider an AI model as less fair if the model is biased against their group. This discrepancy highlights the importance of going beyond user engagement when assessing the fairness of an AI model. In other words, 
Our findings imply that when decision subjects can repeatedly and strategically respond to an AI model, %have the opportunities to improve their qualification and decide whether to keep interacting with an AI model, 
the equality in engagement across different groups of decision subjects should {\em not} be used as a proxy indicator of a fair AI model. Instead, 
responsible AI practitioners should 
delve deeper into truly understanding decision subjects' perceptions of and satisfaction with the AI model to develop fair AI-based decision systems. On the other hand, compared to previous findings~\cite{altugPaper, haiyiPaper}, results of our study highlight that providing decision subjects with opportunities to improve their qualification can effectively maintain user retention in the case that the AI model exhibits a degree of unfairness across groups, especially for subjects whose group is placed at the disadvantaged position by the AI model. This means that when fairness concerns have been identified in AI-based decision systems, %institutions that employ AI-based decision systems (e.g., banks) identify fairness concerns of their AI models, 
actively providing decision subjects with information and guidance on qualification improvement (e.g., via providing algorithmic recourse plans) might be used as a temporary solution to maintain user retention. This may buy the system developers some time to address the fairness concerns of the AI models without losing a significant sector of users. However, we emphasize that this short-term relief should not replace more fundamental and comprehensive solutions which address the root cause fairness problems of the AI model and truly improve decision subjects' satisfactions with the model. In fact, in a real-world environment with competitions (e.g., multiple banks providing loan lending services), it may be users' fairness perceptions and satisfaction that play the key role in shaping their long-term retention.

\vspace{2pt}
\noindent \textbf{Limitations and future work.} 
%There are a few limitations of our study. \erase{First,} We acknowledge that decision subjects may not always want to or have the option to strategically respond to the AI model's decisions on them. \erase{For example, when the AI model was used for disease diagnosis, decision subjects may not be motivated to strategically respond to AI's decisions on them.} However, when a decision involves allocating desirable resources (e.g., loans, jobs, grants) and there exists some measurements of ``qualification'' for such decision, decision subjects may want to and have means to strategically respond to AI's decisions on them, and our study aims to capture such decision domain. 
We acknowledge that our findings were constrained by our study setup. For example, the ways that subjects were incentivized in our study, albeit designed to reflect the uncertainty, gains, and losses that they would encounter in the real-world loan lending scenarios, may have nudged our subjects into more rational thinking and made them put a higher weight on utility-related considerations when determining how to engage with the AI model. The subject population that we engaged through the online platform (i.e., MTurk) may also lean more towards rational thinking, as one of their primary goals is to maximize their earnings. 
Our study was also constrained by a few ``parameters'' of our experimental tasks (e.g., cost for qualification improvement attempt, reward for favorable decision). %For example, \erase{the AI model's loan approval rate changes linearly with applicants' credit levels in our experiment. Thus, findings of our study may reflect decision subjects' reaction to and perceptions of a linear AI model, but may not generalize to other settings where the AI model is non-linear. Other parameters of our experimental tasks include} 
%the cost we define for each qualification improvement attempt, the cost for each loan application, the reward for each favorable decision, and the linear qualification improvement schemes we adopt in our studies. 
Future research could systematically vary these ``parameters'' and investigate how they affect decision subjects' behavior and perceptions. Additionally, the AI model used in our experiment is not updating over time as new data gets generated from decision subjects' interactions with the model.
Empirically investigating the feedback loop between AI model updates and user reactions over time could be another critical direction to explore in the future for advancing our understanding of the long-term impact of AI fairness. 

%% file: Appendix.tex
\label{sec:appendix}

\section{Example Flowchart}

\begin{figure*}[t]
    \centering  \includegraphics[width=0.7\linewidth]{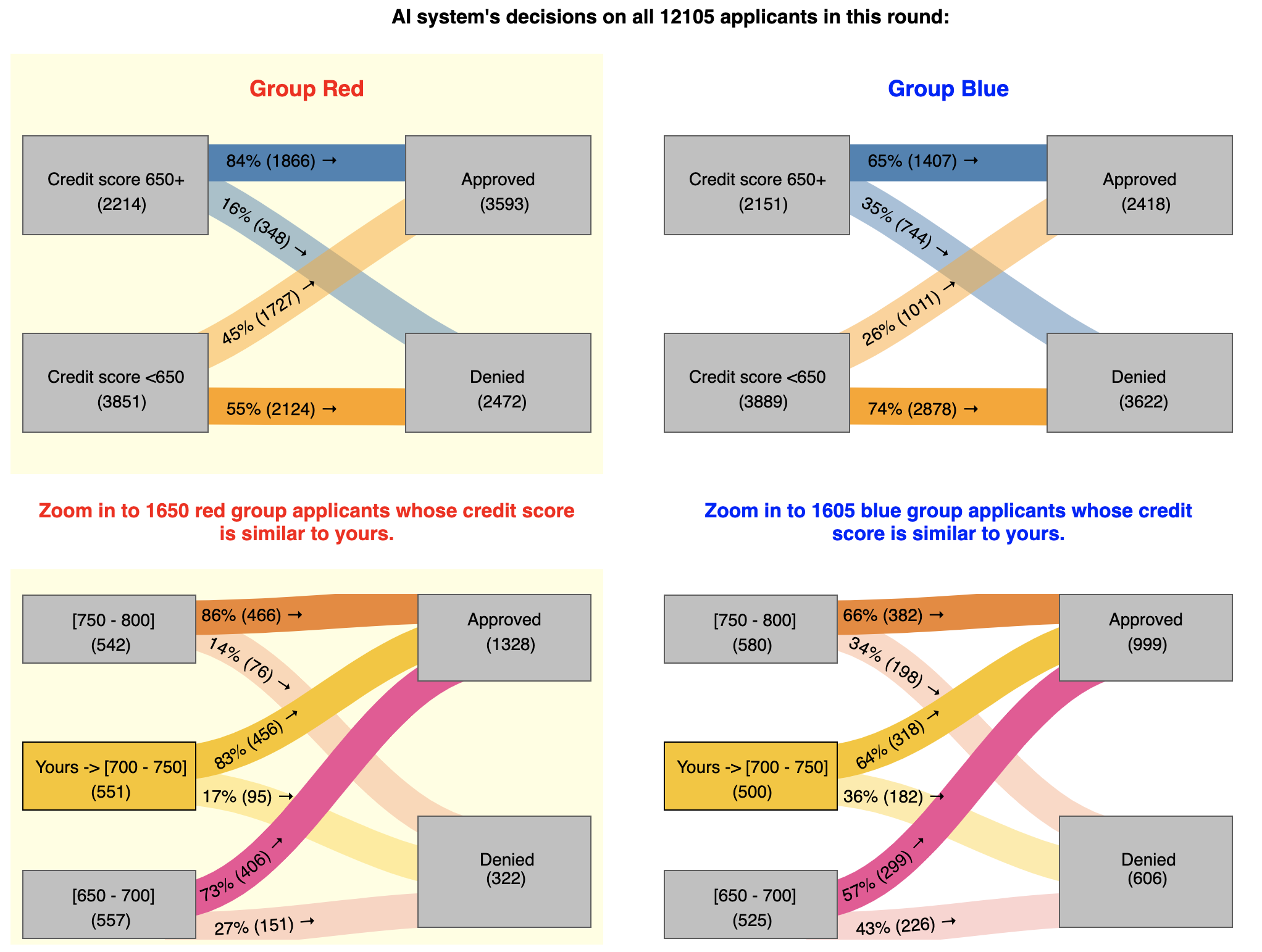}
    \caption{An example of the flowchart that subjects in the unfair AI treatment saw in the experiment, which summarizes the AI model's decisions on different groups of applicants in the past round. Subjects could see the frequency of the AI model approving/denying loans %These flowcharts depict the AI model's approve/deny decisions based on credit scores. Specifically, the first set focuses on decisions 
    both for applicants with/without ``high'' credit scores (i.e., a score of at least 650), and for applicants with similar credit scores as themselves (i.e., applicants with the same credit scores as themselves or one level above/below themselves).  
    %while the second set provides a detailed view for applicants whose credit level is relative to the subject: one level higher, equivalent, or one level lower.
    }
    \label{fig:flowcharts}
    \vspace{-5pt}
\end{figure*}

Figure~\ref{fig:flowcharts} shows an example flowchart that participants saw in our experiment, which illustrates the AI model's decisions on loan applicants of different groups in one round.

\section{A Complete List of Questions Used in the Post-Experiment Survey}\label{sec:survey}

In the post-experiment survey, the questions listed below were used. Please note that for any negative statements, we reversed the rating when calculating the scores for specific measurements, such as risk attitude.

\begin{enumerate}
    \item \textbf{DEMOGRAPHICS QUESTIONS}

    \begin{itemize}
        \item \textbf{What is your age?}
        \begin{enumerate}[label=(\alph*)]
            \item 18-24 years old
            \item 25-34 years old
            \item 35-44 years old
            \item 45-54 years old
            \item 55-64 years old
            \item 65-74 years old
            \item 75 years or older
        \end{enumerate}

        \item \textbf{Which race or ethnicity best describes you? (Please choose only one.)}
        \begin{enumerate}[label=(\alph*)]
            \item American Indian or Alaskan Native
            \item Asian / Pacific Islander
            \item Black or African American
            \item Hispanic
            \item White / Caucasian
            \item Multiple ethnicity/ Other
        \end{enumerate}

        \item \textbf{Please select the state you are in:}
        \begin{enumerate}[label=(\alph*)]
            \item Alabama
            \item Alaska
            \item Arizona
            \item ... 
        \end{enumerate}

        \item \textbf{In general, would you describe your political party of affiliation as:}
        \begin{enumerate}[label=(\alph*)]
            \item Democrat
            \item Republican
            \item Independent
        \end{enumerate}

        \item \textbf{In general, would you describe your political view as:}
        \begin{enumerate}[label=(\alph*)]
            \item Very liberal
            \item Liberal
            \item Somewhat liberal
            \item Moderate
            \item Somewhat conservative
            \item Conservative
            \item Very conservative
        \end{enumerate}

        \item \textbf{What is the highest degree or level of school you have completed (if currently enrolled, highest degree received)?}
        \begin{enumerate}[label=(\alph*)]
            \item No schooling completed
            \item Nursery school to 8th grade
            \item Some high school, no diploma
            \item High school graduate, diploma or the equivalent (for example: GED)
            \item Some college credit, no degree
            \item Trade/technical/vocational training
            \item Associate degree
            \item Bachelor’s degree
            \item Master’s degree
            \item Professional degree
            \item Doctorate degree
        \end{enumerate}
    \end{itemize}
    
    \item \textbf{Sensitivity to Fairness Questions}

    For the rest of the statements below about you (until the end of the survey), please indicate how much you agree with it.

    \begin{itemize}
        \item \textbf{It is very important to me that an AI system making decisions about people is fair (i.e., it treats everyone fairly and does not discriminate). (Positive)}
        \begin{enumerate}[label=(\alph*)]
            \item Strongly disagree
            \item Disagree
            \item Neutral
            \item Agree
            \item Strongly agree
        \end{enumerate}

        \item \textbf{I would only use an AI system if it is fair to everyone. (Positive)}
        \begin{enumerate}[label=(\alph*)]
            \item Strongly disagree
            \item Disagree
            \item Neutral
            \item Agree
            \item Strongly agree
        \end{enumerate}

        \item \textbf{I would stop using an AI system if it is unfair, even if it tends to be in favor of me. (Positive)}
        \begin{enumerate}[label=(\alph*)]
            \item Strongly disagree
            \item Disagree
            \item Neutral
            \item Agree
            \item Strongly agree
        \end{enumerate}

        \item \textbf{When I decide whether to use an AI system or not, I seldom think about whether the system is fair. (Negative)}
        \begin{enumerate}[label=(\alph*)]
            \item Strongly disagree
            \item Disagree
            \item Neutral
            \item Agree
            \item Strongly agree
        \end{enumerate}
    \end{itemize}

    \item \textbf{Risk Attitude Questions}
    
    \begin{itemize}
        \item \textbf{If I were betting on horses and were a big winner in the third or fourth race, I would be more likely to stop playing and take my winnings. (Negative)}
        \begin{enumerate}[label=(\alph*)]
            \item Strongly disagree
            \item Disagree
            \item Neutral
            \item Agree
            \item Strongly agree
        \end{enumerate}

        \item \textbf{I like to do frightening things. (Positive)}
        \begin{enumerate}[label=(\alph*)]
            \item Strongly disagree
            \item Disagree
            \item Neutral
            \item Agree
            \item Strongly agree
        \end{enumerate}

        \item \textbf{I like new and exciting experiences, even if I have to break the rules. (Positive)}
        \begin{enumerate}[label=(\alph*)]
            \item Strongly disagree
            \item Disagree
            \item Neutral
            \item Agree
            \item Strongly agree
        \end{enumerate}
        \item \textbf{I prefer friends who are exciting and unpredictable. (Positive)}
        \begin{enumerate}[label=(\alph*)]
            \item Strongly disagree
            \item Disagree
            \item Neutral
            \item Agree
            \item Strongly agree
        \end{enumerate}
        
        \item \textbf{In general, it is very easy for me to accept taking risks. (Positive)}
        \begin{enumerate}[label=(\alph*)]
            \item Strongly disagree
            \item Disagree
            \item Neutral
            \item Agree
            \item Strongly agree
        \end{enumerate}
    \end{itemize}

    \item \textbf{Empathy Questions}

\begin{itemize}
    \item \textbf{I remain unaffected when someone close to me is happy. (Negative)}
    \begin{enumerate}[label=(\alph*)]
        \item Strongly disagree
        \item Disagree
        \item Neutral
        \item Agree
        \item Strongly agree
    \end{enumerate}
    
    \item \textbf{I enjoy making other people feel better. (Positive)}
    \begin{enumerate}[label=(\alph*)]
        \item Strongly disagree
        \item Disagree
        \item Neutral
        \item Agree
        \item Strongly agree
    \end{enumerate}
    
    \item \textbf{I get a strong urge to help when I see someone who is upset. (Positive)}
    \begin{enumerate}[label=(\alph*)]
        \item Strongly disagree
        \item Disagree
        \item Neutral
        \item Agree
        \item Strongly agree
    \end{enumerate}

    \item \textbf{When I see someone being treated unfairly, I do not feel very much pity for them. (Negative)}
    \begin{enumerate}[label=(\alph*)]
        \item Strongly disagree
        \item Disagree
        \item Neutral
        \item Agree
        \item Strongly agree
    \end{enumerate}
    
    \item \textbf{When I see someone being taken advantage of, I feel kind of protective towards them. (Positive)}
    \begin{enumerate}[label=(\alph*)]
        \item Strongly disagree
        \item Disagree
        \item Neutral
        \item Agree
        \item Strongly agree
    \end{enumerate}
\end{itemize}

\item \textbf{Perceptions of AI Fairness Questions}

\begin{itemize}
    \item \textbf{The bank's AI system is fair. (Positive)}
    \begin{enumerate}[label=(\alph*)]
        \item Strongly disagree
        \item Disagree
        \item Neutral
        \item Agree
        \item Strongly agree
    \end{enumerate}
    
    \item \textbf{The bank's AI system is fair to loan applicants. (Positive)}
    \begin{enumerate}[label=(\alph*)]
        \item Strongly disagree
        \item Disagree
        \item Neutral
        \item Agree
        \item Strongly agree
    \end{enumerate}
    
    \item \textbf{The bank's AI system is fair to manage loan applications. (Positive)}
    \begin{enumerate}[label=(\alph*)]
        \item Strongly disagree
        \item Disagree
        \item Neutral
        \item Agree
        \item Strongly agree
    \end{enumerate}
    
    \item \textbf{The decisions that the bank makes as a result of the AI system will be fair. (Positive)}
    \begin{enumerate}[label=(\alph*)]
        \item Strongly disagree
        \item Disagree
        \item Neutral
        \item Agree
        \item Strongly agree
    \end{enumerate}
    
    \item \textbf{The bank's AI system will lead the bank to make great loan lending decisions. (Positive)}
    \begin{enumerate}[label=(\alph*)]
        \item Strongly disagree
        \item Disagree
        \item Neutral
        \item Agree
        \item Strongly agree
    \end{enumerate}
    
    \item \textbf{The bank's AI system will make mistakes. (Negative)}
    \begin{enumerate}[label=(\alph*)]
        \item Strongly disagree
        \item Disagree
        \item Neutral
        \item Agree
        \item Strongly agree
    \end{enumerate}

    \item \textbf{Free-form text:} Based on your responses to the survey questions above, it seems like that you feel that in this game, the bank's AI-powered system is \textbf{X} to loan applicants in general. If this is correct, please briefly explain why you believe so; if this is incorrect, please briefly describe what do you think of the bank's AI system in terms of its level of fairness. (Add up the ratings for the ``Perceptions of AI Fairness Questions.'' If the total is less than 15, replace \textbf{X} with ``unfair.'' If the total is equal to or greater than 15, replace \textbf{X} with ``fair.'')
\end{itemize}
    
\end{enumerate}

\section{Filtering Procedures Used in the Experiment}\label{sec:filtering}

To ensure that our experimental data was provided by genuine human subjects rather than bots or spammers, we implemented a few protective procedures. First, we incorporated both Google's reCAPTCHA v3\footnote{\url{https://www.google.com/recaptcha/about/}} and a honeypot CAPTCHA (i.e., a CAPTCHA that is hidden in the HTML, thus invisible to real human subjects but visible to bots) in the web application of our experiment to filter out bots. Second, we included an attention check question in the post-experiment survey, which instructed the subject to select a pre-defined option, to filter out inattentive subjects. Finally, we manually checked the subject's responses to open-ended questions in the survey, and filtered out potential spammers (e.g., subjects who provided identical responses or  responses that were not comprehensible). A subject's data was only considered valid if it can pass all these three filtering procedures.

\section{Participant Demographics}\label{sec:demographic}

\begin{table}[h]
\centering
\begin{adjustbox}{width=0.45\textwidth}
\begin{tabular}{@{}lrrrr@{}}
\toprule
\textbf{\%} & \textbf{Study 1} & \textbf{Study 2 (Easy to hard)} & \textbf{Study 2 (Hard to easy)} & \textbf{Study 3} \\
\midrule
\textbf{Gender} & & & \\
\quad Male & 54\% & 51\% & 57\% & 55\%\\
\quad Female & 45\% & 48\% & 42\% & 45\%\\
\quad Other & 1\% & 1\% & 1\% & 0\%\\
\midrule
\textbf{Age} & & & \\
\quad 18--24 & 17\% & 12\% & 15\% & 15\%\\
\quad 25--34 & 29\% & 26\% & 27\% & 26\%\\
\quad 35--44 & 24\% & 26\% & 25\% & 25\%\\
\quad 45+ & 30\% & 36\% & 33\% & 34\%\\
\midrule
\textbf{Race} & & & \\
\quad White & 76\% & 80\% & 77\% & 82\%\\
\quad Black & 7\% & 6\% & 8\% & 6\%\\
\quad Asian & 8\% & 6\% & 6\% & 6\%\\
\quad Hispanic & 3\% & 3\% & 4\% & 3\%\\
\quad Other & 6\% & 5\% & 5\% & 3\%\\
\bottomrule
\end{tabular}
\end{adjustbox}
\caption{Demographics of the subjects 
%and total number of participants 
in each experiment. 
%The demographic percentages are given for Study 1 and Study 2 experiments below for: 
The total number of subjects in each experiment is: 
Study 1: 368 participants, Study 2 (Easy to hard): 328 participants, Study 2 (Hard to easy): 385 participants, and Study 3: 416 participants.}
\label{tab:demographics}
\end{table}

\clearpage
\newpage

\begin{table*}[t]
\section{Study 1: Data Details}
\label{sec:study1More}
\centering
% This will make the font size smaller than \footnotesize
\begin{adjustbox}{width=1\textwidth}
\begin{tabular}{lcccccccccccc}
\hline
& \multicolumn{3}{c}{\textbf{Unfair Red (88)}} & \multicolumn{3}{c}{\textbf{Unfair Blue (92)}} & \multicolumn{3}{c}{\textbf{Fair Red (92)}} & \multicolumn{3}{c}{\textbf{Fair Blue (96)}} \\
\cline{2-13}
\textbf{Variable} & \textbf{Mean} & \textbf{SD} & \textbf{95\% CI} & \textbf{Mean} & \textbf{SD} & \textbf{95\% CI} & \textbf{Mean} & \textbf{SD} & \textbf{95\% CI} & \textbf{Mean} & \textbf{SD} & \textbf{95\% CI} \\
\hline
Improvement & 4.60 & 3.41 & [3.88, 5.32] & 3.63 & 3.04 & [3.00, 4.26] & 3.79 & 3.41 & [3.09, 4.50] & 4.41 & 3.44 & [3.71, 5.10] \\
Retention & 6.91 & 3.78 & [6.11, 7.71] & 6.26 & 3.66 & [5.50, 7.02] & 6.04 & 3.91 & [5.23, 6.85] & 6.41 & 3.95 & [5.61, 7.21] \\
Fairness Perception & 19.0 & 4.75 & [18.0, 20.0] & 17.0 & 5.81 & [15.84, 18.2] & 18.7 & 5.45 & [17.6, 19.9] & 18.8 & 4.87 & [17.8, 19.8] \\
\hline
\end{tabular}
\end{adjustbox}
\caption{Descriptive Statistics of Subjects by Group (Study 1)}
\label{table:descriptive_stats_detailed}
\end{table*}

\begin{figure*}[h]
\centering
\begin{subfigure}[t]{0.3\textwidth}
\centering
    \includegraphics[width=\textwidth]{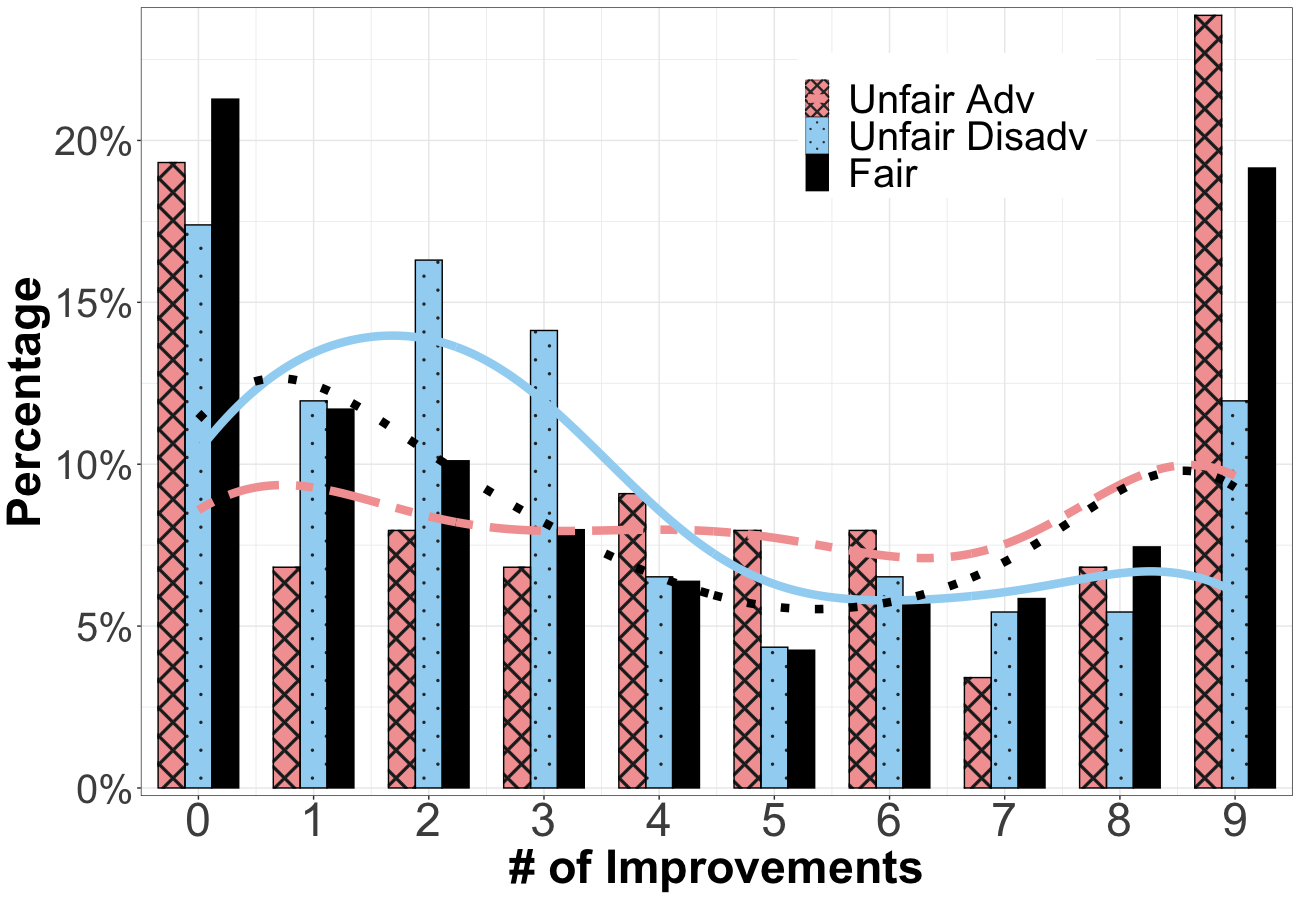}
    \caption{Improvement}
    \label{fig:study1_imp}
\end{subfigure}
%\hfill
\begin{subfigure}[t]{0.3\textwidth} % Reduced width for fitting 4 images in one row
\centering
\includegraphics[width=\textwidth]{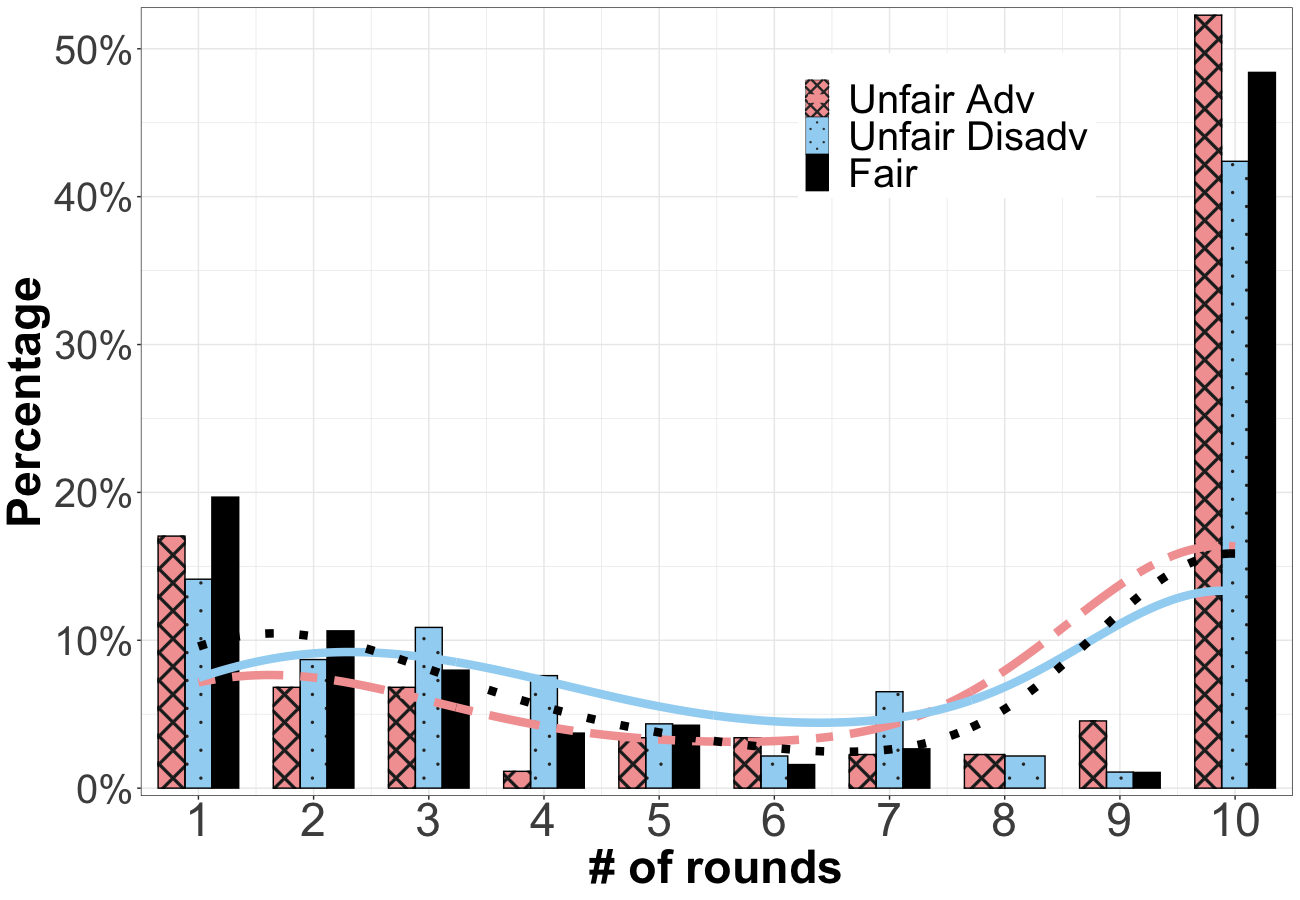}
    \caption{Retention}
    \label{fig:study1_survival}
\end{subfigure}
\caption{Distributions of (a) the number of improvement attempts that subjects made,  and (b) the number of rounds that subjects interacted with the AI model, for subjects who were assigned to the fair AI model treatment, the red (and advantaged) group of the unfair AI model treatment, and the blue (and disadvantaged) group of the unfair AI model treatment, in Study 1. Curves represent the probability density functions obtained through kernel density estimation.}
\label{fig:study1}
\end{figure*}

%\clearpage
%\newpage

\begin{figure*}[h]
\centering
\section{Study 2: Data Details (Easy to hard)}\label{sec:easy2Hard}
\begin{subfigure}[t]{0.3\textwidth}
\centering
    \includegraphics[width=\textwidth]{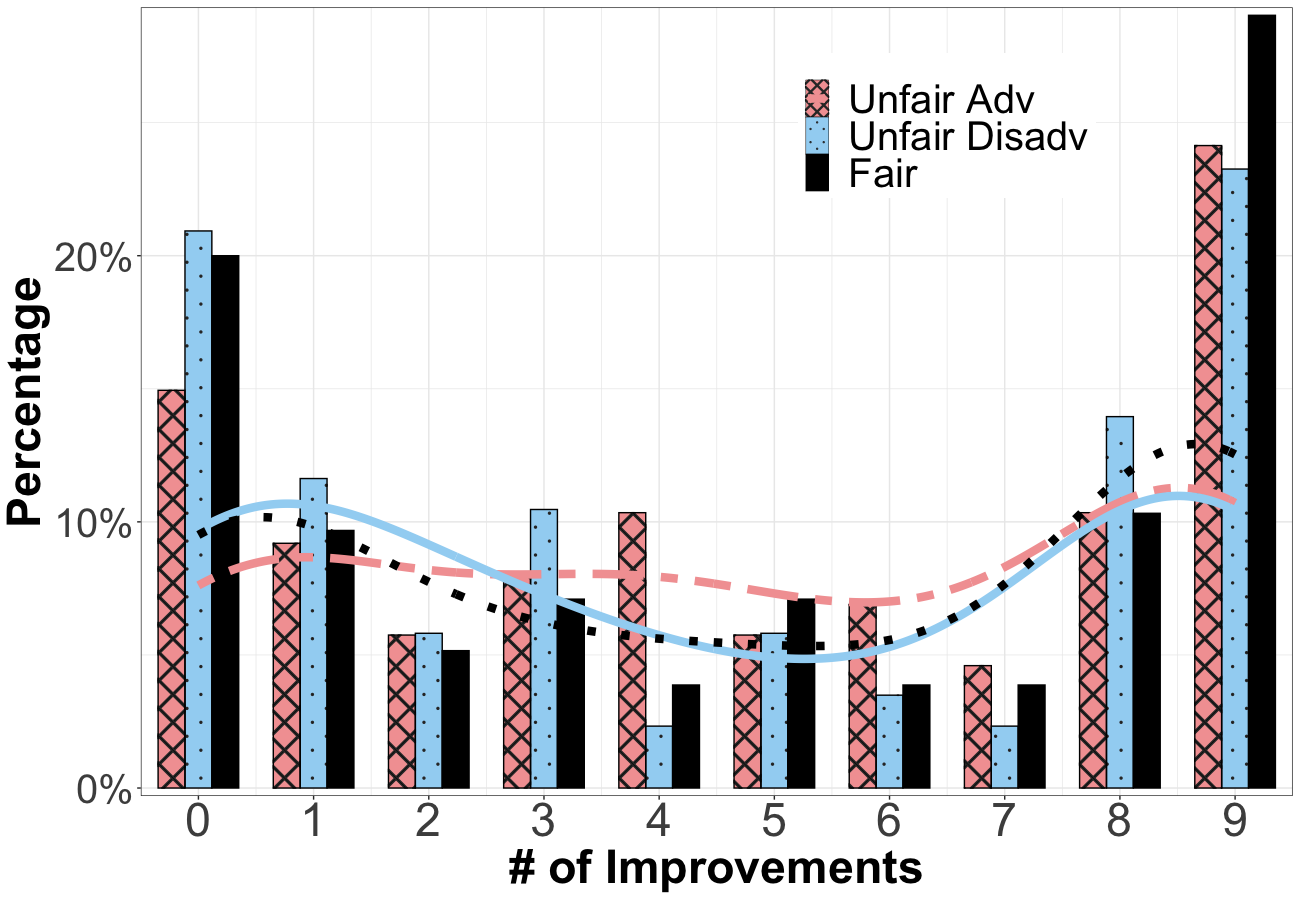}
    \caption{Improvement}
    \label{fig:study2_imp}
\end{subfigure}
%\hfill
\begin{subfigure}[t]{0.3\textwidth} % Reduced width for fitting 4 images in one row
\centering
\includegraphics[width=\textwidth]{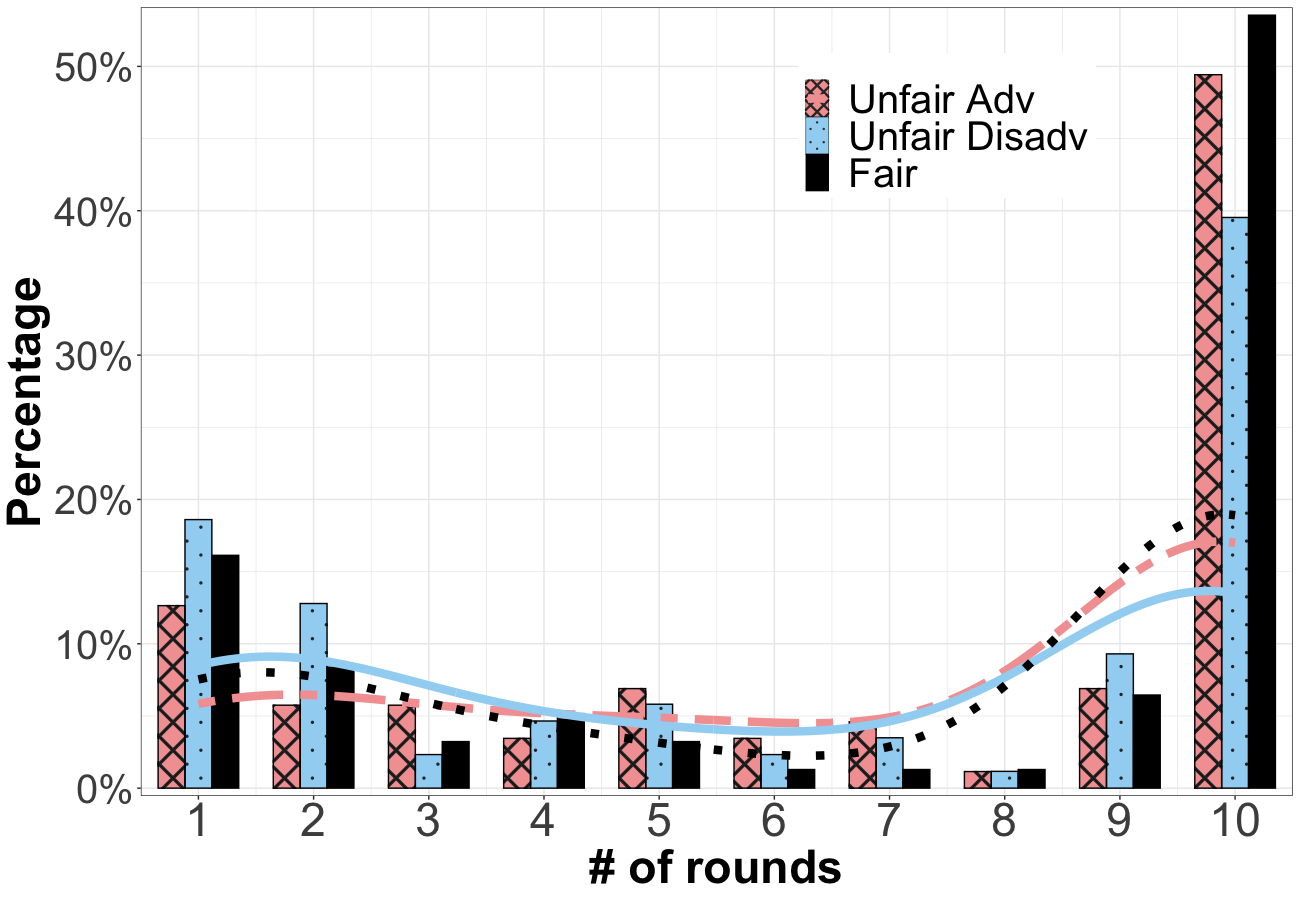}
    \caption{Retention}
    \label{fig:study2_retention}
\end{subfigure}
\caption{Distributions of (a) the number of improvement attempts that subjects made,  and (b) the number of rounds that subjects interacted with the AI model, for subjects who were assigned to the fair AI model treatment, the red (and advantaged) group of the unfair AI model treatment, and the blue (and disadvantaged) group of the unfair AI model treatment, in Study 2 sub-experiment ``Easy to hard''. Curves represent the probability density functions obtained through kernel density estimation.}
\label{fig:combined_study2_e_t_h}
\end{figure*}

\begin{table*}[h]
\centering
%\scriptsize % This will make the font size smaller than \footnotesize
\begin{adjustbox}{width=1\textwidth}
\begin{tabular}{lcccccccccccc}
\hline
& \multicolumn{3}{c}{\textbf{Unfair Red (87)}} & \multicolumn{3}{c}{\textbf{Unfair Blue (86)}} & \multicolumn{3}{c}{\textbf{Fair Red (78)}} & \multicolumn{3}{c}{\textbf{Fair Blue (77)}} \\
\cline{2-13}
\textbf{Variable} & \textbf{Mean} & \textbf{SD} & \textbf{95\% CI} & \textbf{Mean} & \textbf{SD} & \textbf{95\% CI} & \textbf{Mean} & \textbf{SD} & \textbf{95\% CI} & \textbf{Mean} & \textbf{SD} & \textbf{95\% CI} \\
\hline
Improvement & 4.89 & 3.36 & [4.17, 5.60] & 4.51 & 3.62 & [3.74, 5.29] & 5.19 & 3.54 & [4.39, 5.99] & 4.53 & 3.71 & [3.69, 5.37] \\
Retention & 7.08 & 3.52 & [6.33, 7.83] & 6.26 & 3.81 & [5.44, 7.07] & 7.36 & 3.58 & [6.55, 8.17] & 6.64 & 3.94 & [5.74, 7.53] \\
Fairness Perception & 18.6 & 5.25 & [17.5, 19.7] & 18.9 & 4.99 & [17.8, 20.0] & 18.6 & 4.94 & [17.5, 19.7] & 19.5 & 4.45 & [18.5, 20.5] \\
\hline
\end{tabular}
\end{adjustbox}
\caption{Descriptive Statistics of Subjects by Group (Study 2: Easy to hard)}
\label{table:descriptive_stats_detailed_e_to_h}
\end{table*}

\clearpage
\newpage

\begin{figure*}[h]
\centering
\section{Study 2: Data Details (Hard to easy)}\label{sec:hard2Easy}
\begin{subfigure}[t]{0.3\textwidth}
\centering
    \includegraphics[width=\textwidth]{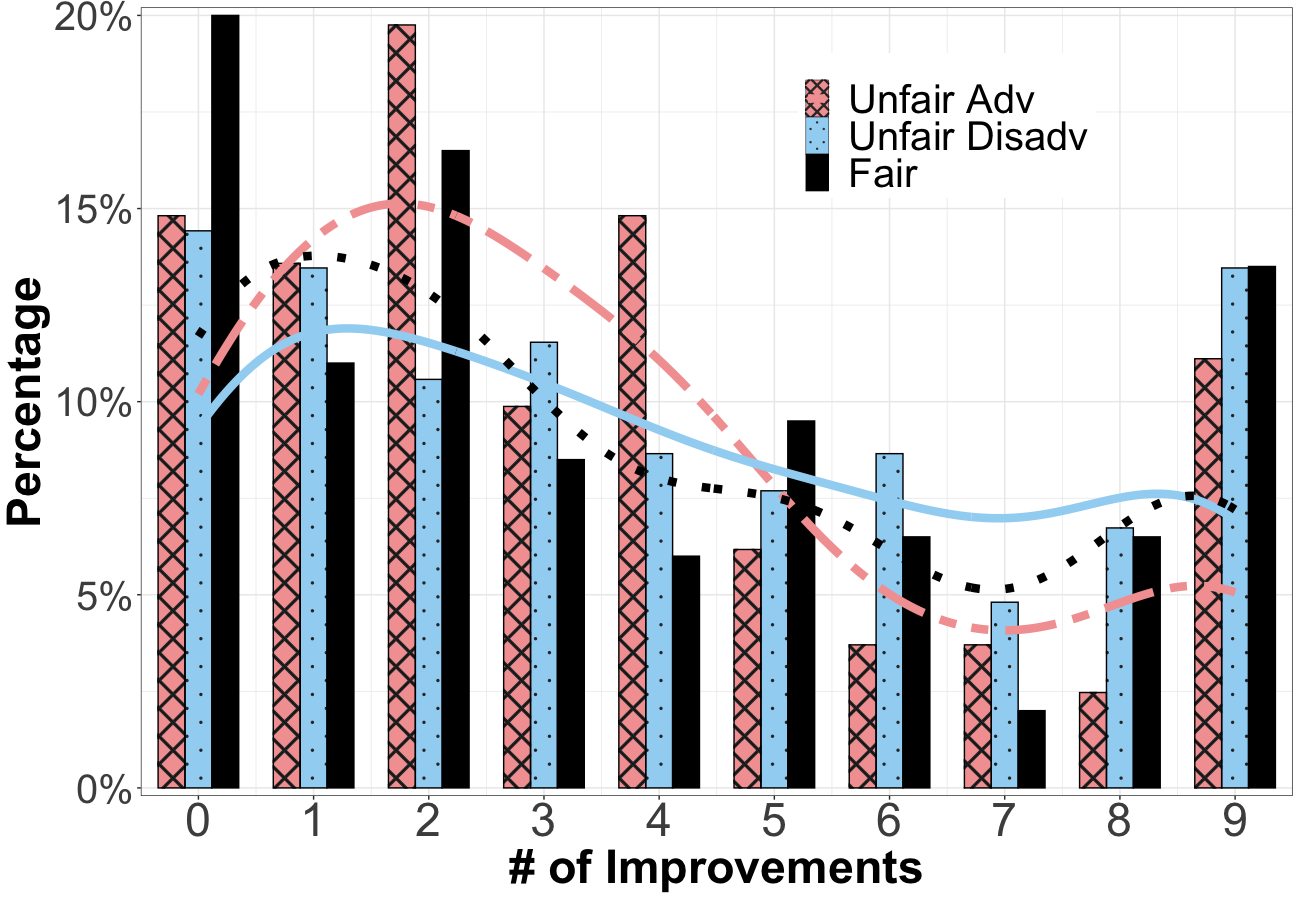}
    \caption{Improvement}
    \label{fig:study2_imp1}
\end{subfigure}
%\hfill
\begin{subfigure}[t]{0.3\textwidth} % Reduced width for fitting 4 images in one row
\centering
\includegraphics[width=\textwidth]{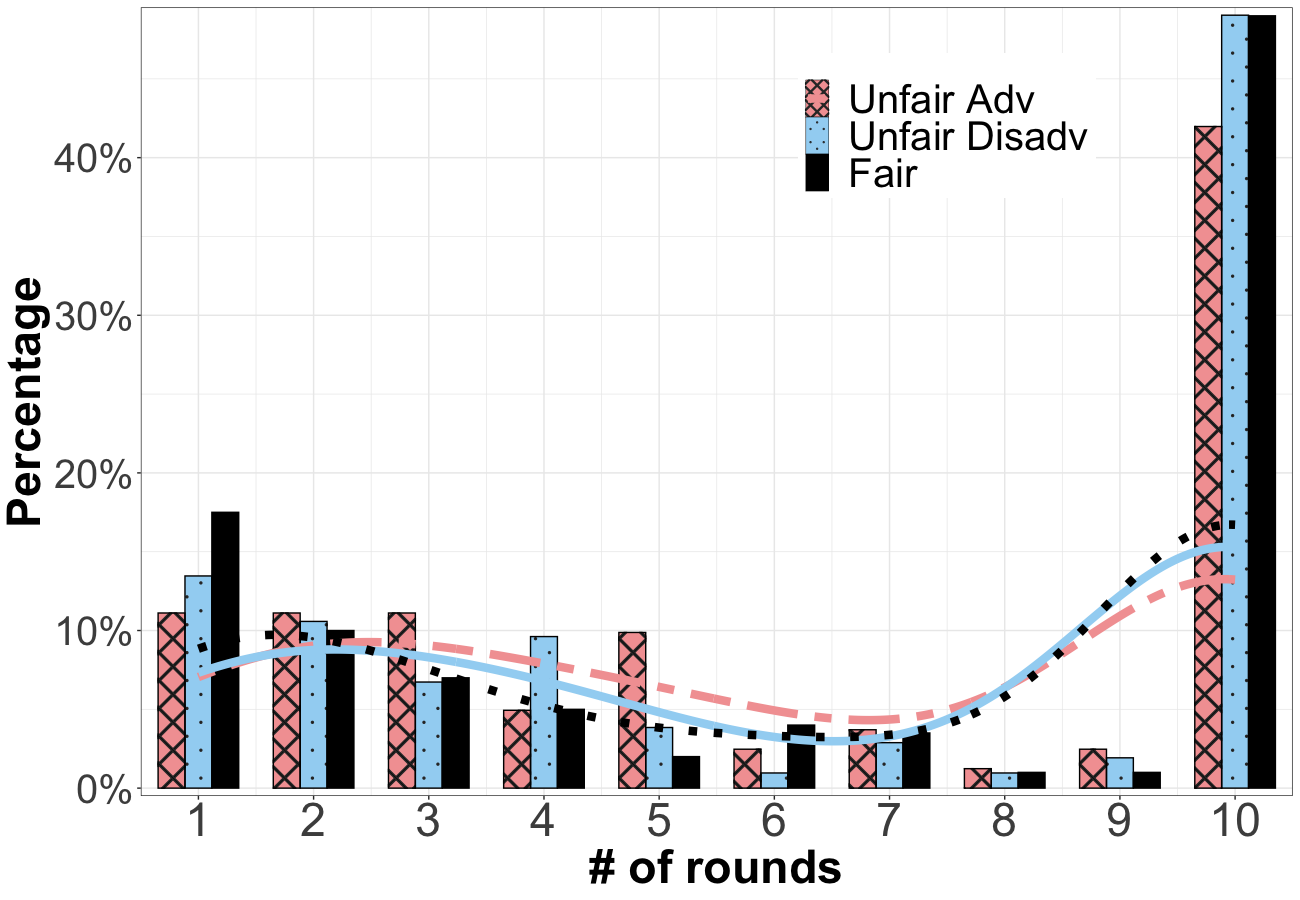}
    \caption{Retention}
    \label{fig:study2_retention1}
\end{subfigure}
\caption{Distributions of (a) the number of improvement attempts that subjects made,  and (b) the number of rounds that subjects interacted with the AI model, for subjects who were assigned to the fair AI model treatment, the red (and advantaged) group of the unfair AI model treatment, and the blue (and disadvantaged) group of the unfair AI model treatment, in Study 2 sub-experiment ``Hard to easy''. Curves represent the probability density functions obtained through kernel density estimation.}
\label{fig:combined_study2_h_t_e}
\end{figure*}

% \newpage

\begin{table*}[h]
\centering
%\scriptsize % This will make the font size smaller than \footnotesize
\begin{adjustbox}{width=1\textwidth}
\begin{tabular}{lcccccccccccc}
\hline
& \multicolumn{3}{c}{\textbf{Unfair Red (81)}} & \multicolumn{3}{c}{\textbf{Unfair Blue (104)}} & \multicolumn{3}{c}{\textbf{Fair Red (96)}} & \multicolumn{3}{c}{\textbf{Fair Blue (104)}} \\
\cline{2-13}
\textbf{Variable} & \textbf{Mean} & \textbf{SD} & \textbf{95\% CI} & \textbf{Mean} & \textbf{SD} & \textbf{95\% CI} & \textbf{Mean} & \textbf{SD} & \textbf{95\% CI} & \textbf{Mean} & \textbf{SD} & \textbf{95\% CI} \\
\hline
Improvement & 3.41 & 2.82 & [2.78, 4.03] & 4.03 & 3.07 & [3.43, 4.63] & 3.26 & 3.07 & [2.64, 3.88] & 4.06 & 3.16 & [3.44, 4.67] \\
Retention & 6.28 & 3.60 & [5.49, 7.08] & 6.54 & 3.74 & [5.81, 7.27] & 6.30 & 3.92 & [5.51, 7.10] & 6.57 & 3.77 & [5.83, 7.30] \\
Fairness Perception & 18.8 & 5.34 & [17.6, 19.9] & 17.3 & 5.75 & [16.1, 18.4] & 19.8 & 4.02 & [19.0, 20.7] & 18.9 & 4.81 & [18.0, 19.9] \\
\hline
\end{tabular}
\end{adjustbox}
\caption{Descriptive Statistics of Subjects by Group (Study 2: Hard to easy)}
\label{table:descriptive_stats_detailed_h_to_e}
\end{table*}

%\clearpage
%\newpage

\begin{figure*}[h]
\centering
\section{Study 3: Data Details}\label{sec:study3}
\begin{subfigure}[t]{0.3\textwidth}
\centering
    \includegraphics[width=\textwidth]{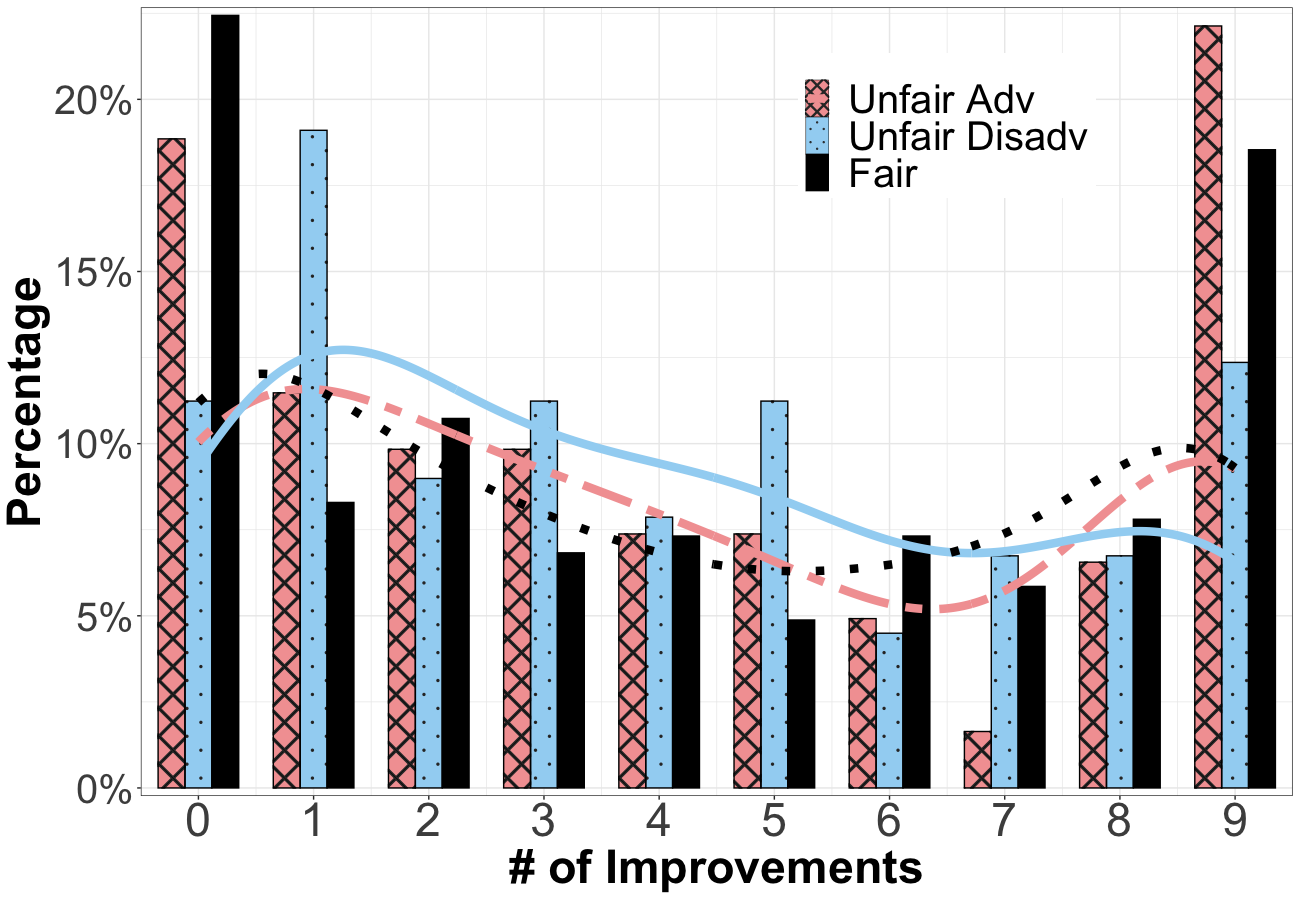}
    \caption{Improvement}
    \label{fig:study3_imp}
\end{subfigure}
%\hfill
\begin{subfigure}[t]{0.3\textwidth} % Reduced width for fitting 4 images in one row
\centering
\includegraphics[width=\textwidth]{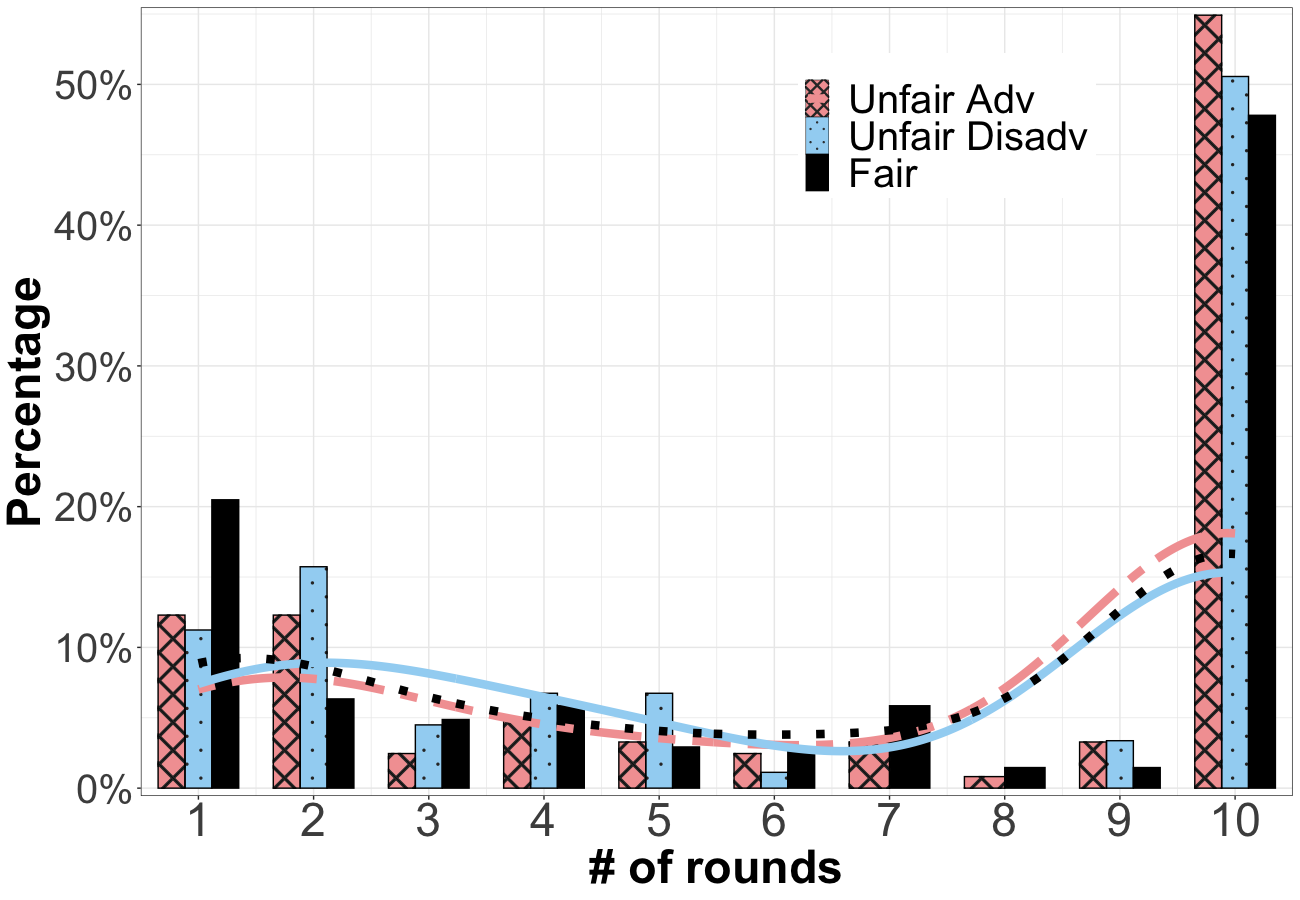}
    \caption{Retention}
    \label{fig:study3_ret}
\end{subfigure}
\caption{Distributions of (a) the number of improvement attempts that subjects made,  and (b) the number of rounds that subjects interacted with the AI model, for subjects who were assigned to the fair AI model treatment, the red (and advantaged) group of the unfair AI model treatment, and the blue (and disadvantaged) group of the unfair AI model treatment, in Study 3. Curves represent the probability density functions obtained through kernel density estimation.}
\label{fig:combined_study3}
\end{figure*}

\begin{table*}[h]
\centering
%\scriptsize % This will make the font size smaller than \footnotesize
\begin{adjustbox}{width=1\textwidth}
\begin{tabular}{lcccccccccccc}
\hline
& \multicolumn{3}{c}{\textbf{Unfair Male (122)}} & \multicolumn{3}{c}{\textbf{Unfair Female (89)}} & \multicolumn{3}{c}{\textbf{Fair Male (107)}} & \multicolumn{3}{c}{\textbf{Fair Female (98)}} \\
\cline{2-13}
\textbf{Variable} & \textbf{Mean} & \textbf{SD} & \textbf{95\% CI} & \textbf{Mean} & \textbf{SD} & \textbf{95\% CI} & \textbf{Mean} & \textbf{SD} & \textbf{95\% CI} & \textbf{Mean} & \textbf{SD} & \textbf{95\% CI} \\
\hline
Improvement & 4.20 & 3.41 & [3.59, 4.81] & 3.98 & 3.03 & [3.34, 4.62] & 4.36 & 3.50 & [3.69, 5.04] & 3.98 & 3.31 & [3.32, 4.64] \\
Retention & 7.03 & 3.70 & [6.37, 7.70] & 6.60 & 3.78 & [5.80, 7.39] & 6.60 & 3.77 & [5.87, 7.32] & 6.34 & 3.88 & [5.56, 7.12] \\
Fairness Perception & 18.9 & 5.64 & [17.9, 19.9] & 17.3 & 5.79 & [16.1, 18.5] & 19.4 & 4.85 & [18.5, 20.3] & 18.4 & 5.79 & [17.2, 19.5] \\
\hline
\end{tabular}
\end{adjustbox}
\caption{Descriptive Statistics of Subjects by Group (Study 3)}
\label{table:descriptive_stats_detailed_gen}
\end{table*}

\clearpage
\newpage

\section{Factors explaining why decision subjects might perceive an AI-based decision system as fair or unfair}\label{sec:fairnessFactors}

To explore factors that drive decision subjects' perceived fairness of an AI model when they can strategically and repeatedly respond to its decisions, we looked into subjects' open-ended text responses in the exit survey, explaining why they perceived the AI model as fair/unfair. Among these responses, we identified evidence suggesting decision subjects' perceived fairness of the AI was influenced by {\em social comparisons}. For example, a subject in the unfair AI treatment believed the AI model as unfair because ``{\em people in red and blue groups with the exact same credit score ranges had different chances of getting approved}''.  Meanwhile, subjects' fairness perceptions of the AI model could also be influenced by {\em temporal self-comparison}. As an example, one subject explained their rationale for perceiving the AI model as fair by stating ``{\em when my credit score improved, I kept getting approved for loans}''.

For many subjects, {\em meritocracy} was a key principle for them to evaluate the AI model's fairness: Observing people with higher credit scores have higher chances of being approved for their loans increased their perceived fairness of the AI model; however,  observing people with similar credit scores get different outcomes or some highly-qualified people get loans denied while some lowly-qualified people get loans approved made them feel that the AI model was inconsistent and unfair. On the other hand, some subjects believed the key principle for fairness should be ``{\em prioritize the needed}'', and they criticized the AI model in the experiment as unfair because ``{\em the AI is geared towards giving approvals to customers who generally don't need loans in the first place (highest credit score demographic)}''. Decision subjects' judgement of the AI's fairness was also influenced by its {\em transparency}. For example, some subject found the AI model to be unfair because ``{\em I was rejected and there was no information given as to why}'', while others commented that they can not definitely assess whether the AI model is fair or unfair because they ``{\em don't understand the internal workings of the AI system well enough}''.

Interestingly, there also exist some influencing factors of subjects' perceived fairness of AI wherein different individuals may hold different opinions. One such factor is the {\em feature} that the AI model uses to make its decisions. For example, some subjects perceived the AI model in the experiment as fair because they believe ``{\em credit scores give an accurate assessment of the likelihood that someone will pay their loans back and it's fair to use those scores when determining credit worthiness}''. However, some other subjects considered the AI model as unfair because they thought AI's focus on credit scores implied it failed to ``{\em take many more factors into account}'' and credit score itself was ``{\em an inherently unfair system}''. Similarly, the fact that AI is pre-programmed is interpreted by different subjects differently---some considered this as an indicator of fairness because ``{\em emotion and bias appears to be left out of AI decision making}'', while others believed it leads to unfair decisions because it means ``{\em AI is not capable of compassion}'' and ``{\em treating people as numbers is not treating them fairly}''. 

\section{Simulation for Parameter Estimation}\label{sec:simulation}
To determine the ideal cost of qualification improvement to be used in our experiment, 
we conduct a simulation by modeling subjects' decision-making process in our experiment as a Markov Decision Process (MDP). Specifically, in our MDP, we identify the following key components: 
%This approach helps us identify the ``ideal'' actions that the subjects could take, considering the specific task design and parameter values. The MDP framework allows us to simulate potential outcomes based on various actions taken by the subjects within the given task context. For our MDP, we identify the following key components:

\begin{itemize}
    \item \textbf{States (s):} There are 13 possible states in total---12 ``regular'' states ($s_1, s_2, \ldots, s_{12}$), each corresponding to one  credit score level (e.g., $s_1$ corresponds to 300--350), and an additional terminal state ($s_0$) representing an exit from the game. At any time $t$, the subject is in one of these 13 states.
    \item \textbf{Actions (a):} In a regular state, there are three actions that a subject could take:
    \begin{itemize}
        \item $A1$: Apply and attempt to improve credit score. (Not available for the highest credit level state $s_{12}$)
        \item $A2$: Apply without attempting to improve credit score.
        \item $A3$: Do not apply and exit the game.
    \end{itemize}
    When the subject is in the terminal state ($s_0$), no action is available to them.
    \item \textbf{Transition Probabilities ($T(s, a, s')$):}
    \begin{itemize}
        \item When $a = A1$, $T(s, a, s') = P(s'|s, a) = 1-p_s$ if $s' = s$ and $T(s, a, s') = p_s$ if $s'$ is the level above $s$; $T(s, a, s') = 0$ for all other $s'$. This is to reflect that the success rate for a subject of credit level $s$ to progress to the next credit level is $p_s$ (e.g., $p_s=0.44$ for all $s$ in Study 1 and 3).  
        \item When $a = A2$, $T(s, a, s') = P(s'|s, a) = 1$ if $s' = s$; $T(s, a, s') = 0$ for all other $s'$.
        \item When $a = A3$, $T(s, a, s') = P(s'|s, a) = 1$ if $s' = s_0$; $T(s, a, s') = 0$ for all other $s'$.
    \end{itemize}
    \item \textbf{Reward Function ($R(s, a, s')$):}
    \begin{itemize}
        \item For $a = A1$, $R(s, a, s') = q_{s'} \cdot 50 - (1 - q_{s'}) \cdot 50 - x$ when $s' = s$ or $s'$ is the level above $s$ ($q_{s'}$ is the probability of getting loan approval when the credit score is at level $s'$, $x$ is the cost for improvement). This is to reflect that when the loan is approved, the subject's net profit is 50 (earn 100 coins but paid 50 coins for application), but when the loan is rejected, the subject's net profit is -50 (paid 50 coins for application). Regardless of whether the loan gets approved, $x$ coins have been paid for qualification improvement.
        \item For $a = A2$, $R(s, a, s') = q_{s} \cdot 50 - (1 - q_{s}) \cdot 50$ (since $s' = s$ for certain).
        \item For $a = A3$, $R(s, a, s') = 0$ (since $s' = s_0$ for certain).
    \end{itemize}
\end{itemize}

%Addressing the decision-making process in loan applications, we utilized a MDP to simulate an ``average’’ subjects' choices. We modeled each credit score level as a distinct state ($s_1, s_2, \ldots, s_{12}$), with a terminal state ($s_0$) representing an exit from the game. Subjects could either attempt to improve their credit score and apply for a loan (A1), apply without improvement (A2), or exit the game (A3). The transition probabilities were defined such that with A1, there was a 56\% chance of remaining at the same score and a 44\% chance of improving to the next level. For A2, the probability was fixed at 100\% to remain the current score, and for A3, it was a certain transition to the terminal state.

%The reward function was designed to balance the expected benefits of loan approval against the costs of applying and potentially improving credit scores. For action A1, the reward was calculated as the loan approval probability at the next credit score level times 50 (50 coins net gain if the loan approved), minus the denial probability times 50 (reflecting the 50 coins fee if denied), adjusted by the cost for improvement. For A2, the reward was the loan approval probability at the current score level times 50, minus the loan denial probability times 50, and for A3, the reward was set to zero, signifying the end of the game.

We considered a set of candidate cost of qualification improvement, i.e., $x\in\{1, 5, 10, 20\}$. For each cost level, we used the value iteration algorithm to solve the corresponding MDP to determine the optimal policy for decision subjects who were interacted with the fair AI model, decision subjects who were interacted with the unfair AI model and were favored by AI, and decision subjects who were interacted with the unfair AI model and were disfavored by AI. Then, to understand the average engagement strategy that a decision subject with an initial credit level of $s\in\{s_1, \cdots, s_9\}$ would take,  
%to understand the expected number of qualification improvements and the expected number of rounds they would apply for loans should they follow the optimal policy, 
we simulated $N=1000$ decision subjects who started from $s$, and took actions based on the optimal policy for a maximum of 9 rounds (as subjects in our experiments were required to apply for a loan in their first round of interaction with AI, and then can decide to continue interacting with the AI model for up to 9 more rounds). We then averaged across these $1000$ simulated subjects to compute the expected number of qualification improvements and the expected number of rounds they would apply for loans for a subject starting from state $s$, and we visualized how these expectations change with the subject's initial credit level.
Moreover, we repeated this simulation for three different qualification improvement schemes: the constant scheme used in Study 1 and 3, the easy-to-hard scheme used in the ``Easy to hard'' sub-experiment in Study 2, and the hard-to-easy scheme used in the ``Hard to easy'' sub-experiment in Study 2.

\begin{figure*}[p]
    \centering
    \vspace*{\fill} % Add vertical space above
    %\subsection*{\centering 1---Coins Improvement Cost Analysis}
    \includegraphics[width=0.75\textwidth]{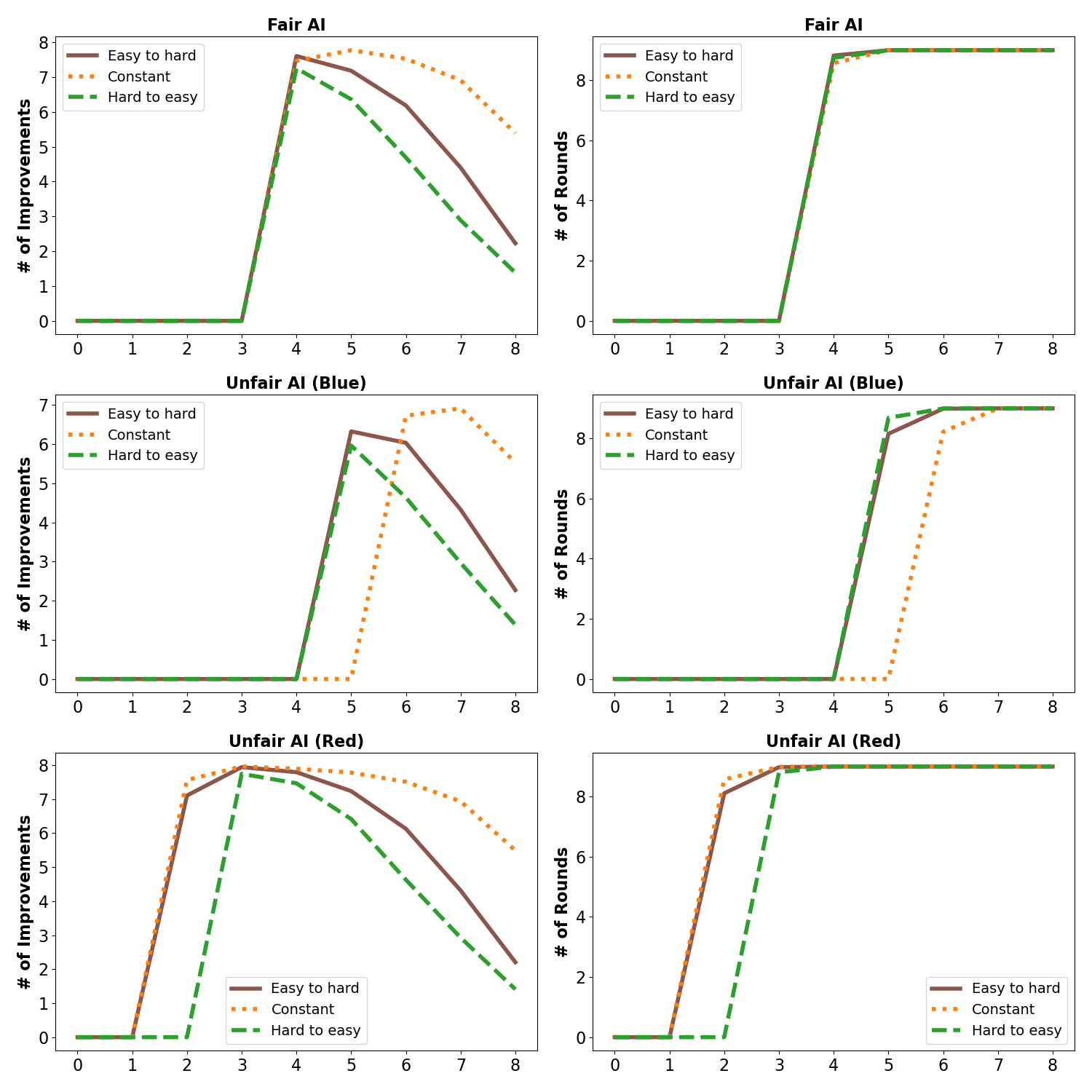}
    \caption{This figure presents the outcomes of implementing a 1-coin improvement cost, considering both Fair and Unfair AI treatments across three studies. The x-axis represents the initial credit score ranges, categorized from 0 to 8, with higher scores indicating higher ranges. In the Fair AI scenario (first row), the figure displays two graphs: the left graph indicates the expected number of improvements, while the right graph shows the expected number of rounds for subjects to keep interacting with the AI model. These are given within each credit score range and for each improvement difficulty used across our three studies. The middle and third rows represent the Unfair AI treatment for blue/female (disadvantaged) and red/male (advantaged) group subjects, similarly displaying ``number of times to improve'' and ``number of rounds to apply for a loan'' strategies as in the first row. The various improvement difficulties are color-coded: constant difficulty used in Studies 1 and 3 (orange), hard to easy in Study 2's sub-experiment (green), and easy to hard in Study 2's other sub-experiment (brown), illustrating different expected strategies for each improvement difficulty across credit score ranges. A low improvement cost might cause most decision subjects to improve since it is easily accessible.}
    \label{fig:1coins}
    \vspace*{\fill} % Add vertical space below
\end{figure*}

\begin{figure*}[p]
    \centering
    \vspace*{\fill} % Add vertical space above
    %\subsection*{\centering 5---Coins Improvement Cost Analysis}
    \includegraphics[width=0.75\textwidth]{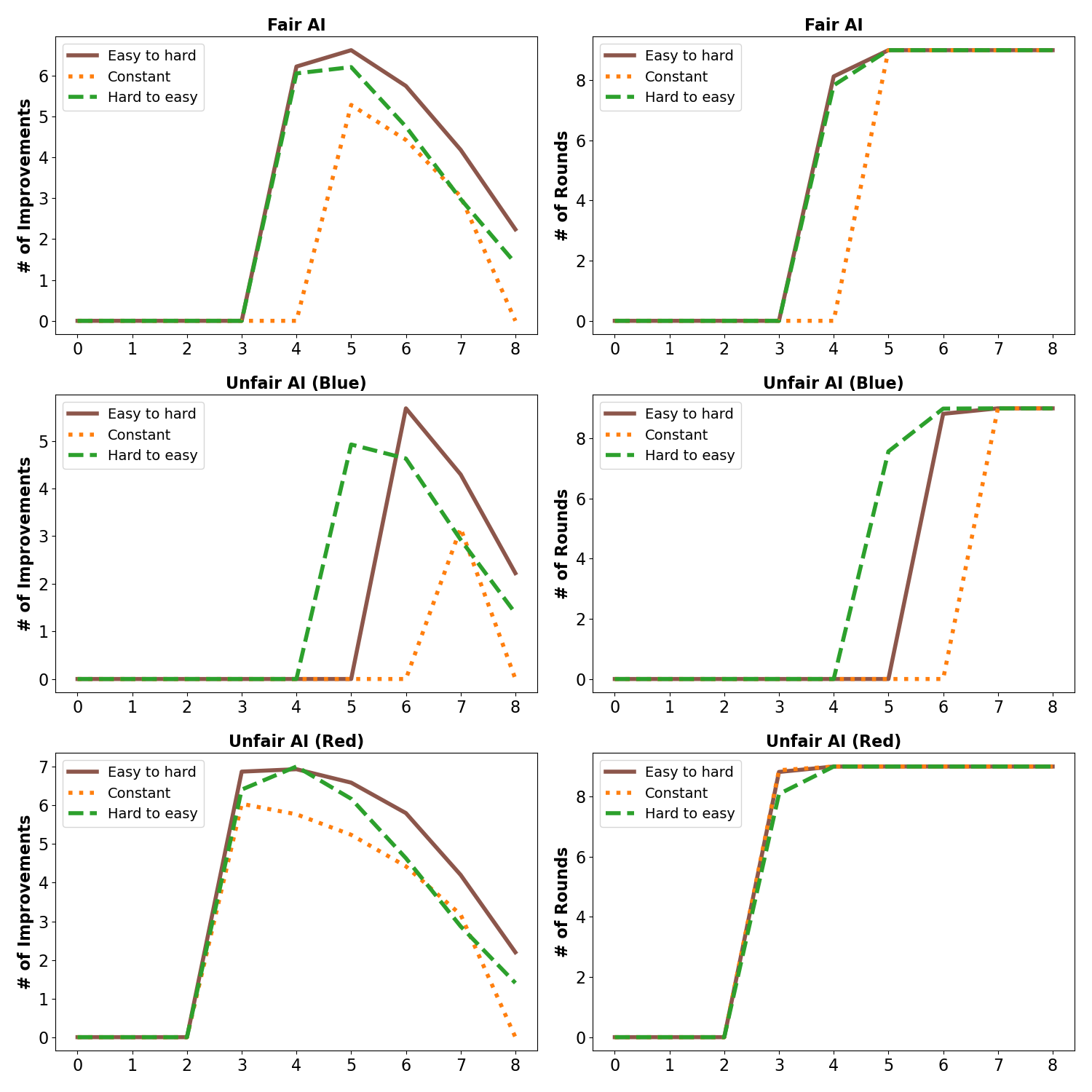}
    \caption{This figure presents the outcomes of implementing a 5-coins improvement cost, considering both Fair and Unfair AI treatments across three studies. The x-axis represents the initial credit score ranges, categorized from 0 to 8, with higher scores indicating higher ranges. In the Fair AI scenario (first row), the figure displays two graphs: the left graph indicates the expected number of improvements, while the right graph shows the expected number of rounds for subjects to keep interacting with the AI model. These are given within each credit score range and for each improvement difficulty used across our three studies. The middle and third rows represent the Unfair AI treatment for blue/female (disadvantaged) and red/male (advantaged) group subjects, similarly displaying ``number of times to improve'' and ``number of rounds to apply for a loan'' strategies as in the first row. The various improvement difficulties are color-coded: constant difficulty used in Studies 1 and 3 (orange), hard to easy in Study 2's sub-experiment (green), and easy to hard in Study 2's other sub-experiment (brown), illustrating different expected strategies for each improvement difficulty across credit score ranges. 5 coins improvement cost seems to be a better fit for our setting. }
    \label{fig:5coins}
    \vspace*{\fill} % Add vertical space below
\end{figure*}

\begin{figure*}[p]
    \centering
    \vspace*{\fill}
    %\subsection*{\centering 10---Coins Improvement Cost Analysis}
    \includegraphics[width=0.75\textwidth]{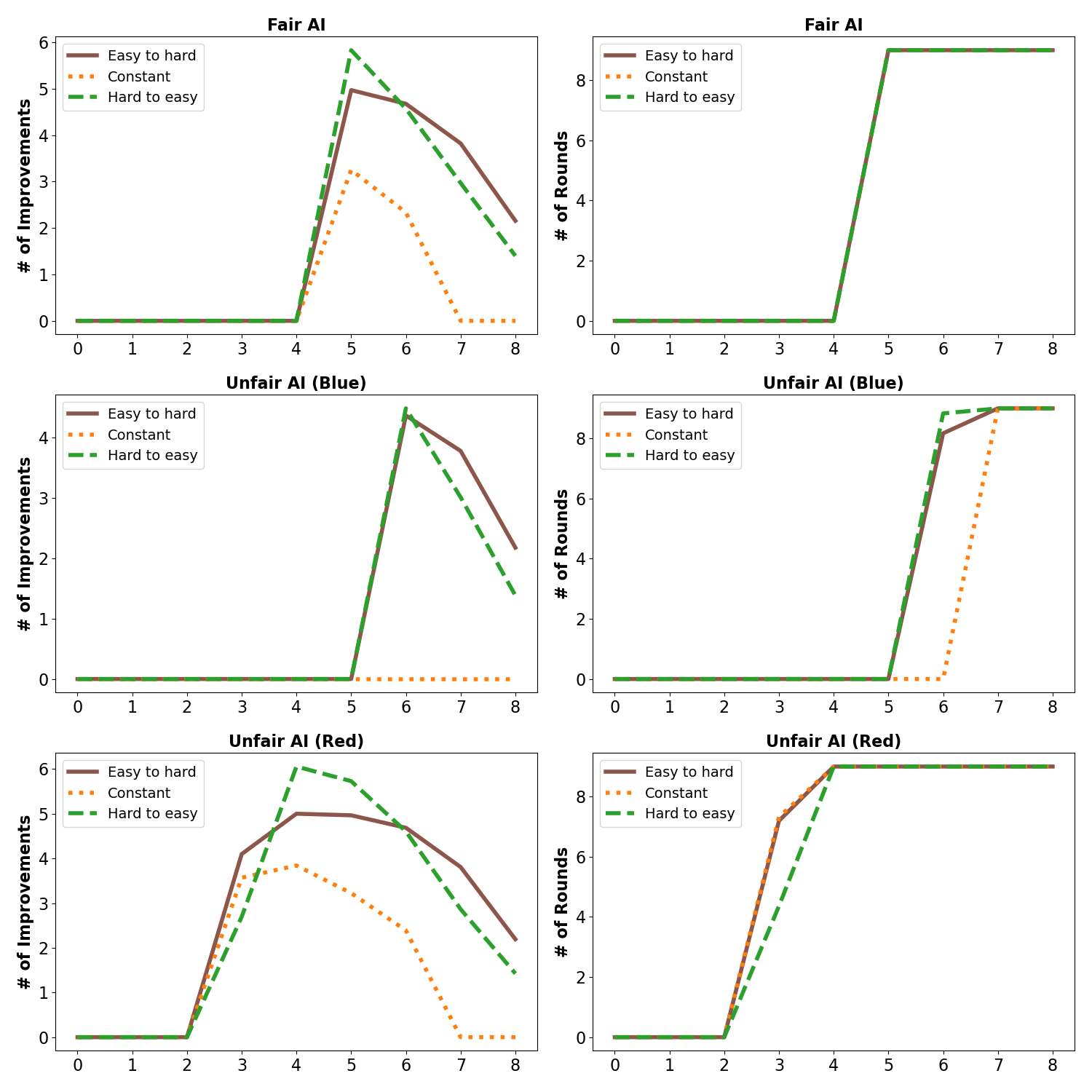}
    \caption{This figure presents the outcomes of implementing a 10-coins improvement cost, considering both Fair and Unfair AI treatments across three studies. The x-axis represents the initial credit score ranges, categorized from 0 to 8, with higher scores indicating higher ranges. In the Fair AI scenario (first row), the figure displays two graphs: the left graph indicates the expected number of improvements, while the right graph shows the expected number of rounds for subjects to keep interacting with the AI model. These are given within each credit score range and for each improvement difficulty used across our three studies. The middle and third rows represent the Unfair AI treatment for blue/female (disadvantaged) and red/male (advantaged) group subjects, similarly displaying ``number of times to improve'' and ``number of rounds to apply for a loan'' strategies as in the first row. The various improvement difficulties are color-coded: constant difficulty used in Studies 1 and 3 (orange), hard to easy in Study 2's sub-experiment (green), and easy to hard in Study 2's other sub-experiment (brown), illustrating different expected strategies for each improvement difficulty across credit score ranges. A high improvement cost might cause most decision subjects to not improve.
    %5 coins improvement cost seems to be a better fit for our setting. The figure highlights the convergence of strategies as the improvement cost increases.
    }
    \label{fig:10coins}
    \vspace*{\fill}
\end{figure*}

\begin{figure*}[h]
    \centering
    \vspace*{\fill}
    %\subsection*{\centering 20---Coins Improvement Cost Analysis}
    \includegraphics[width=0.75\textwidth]{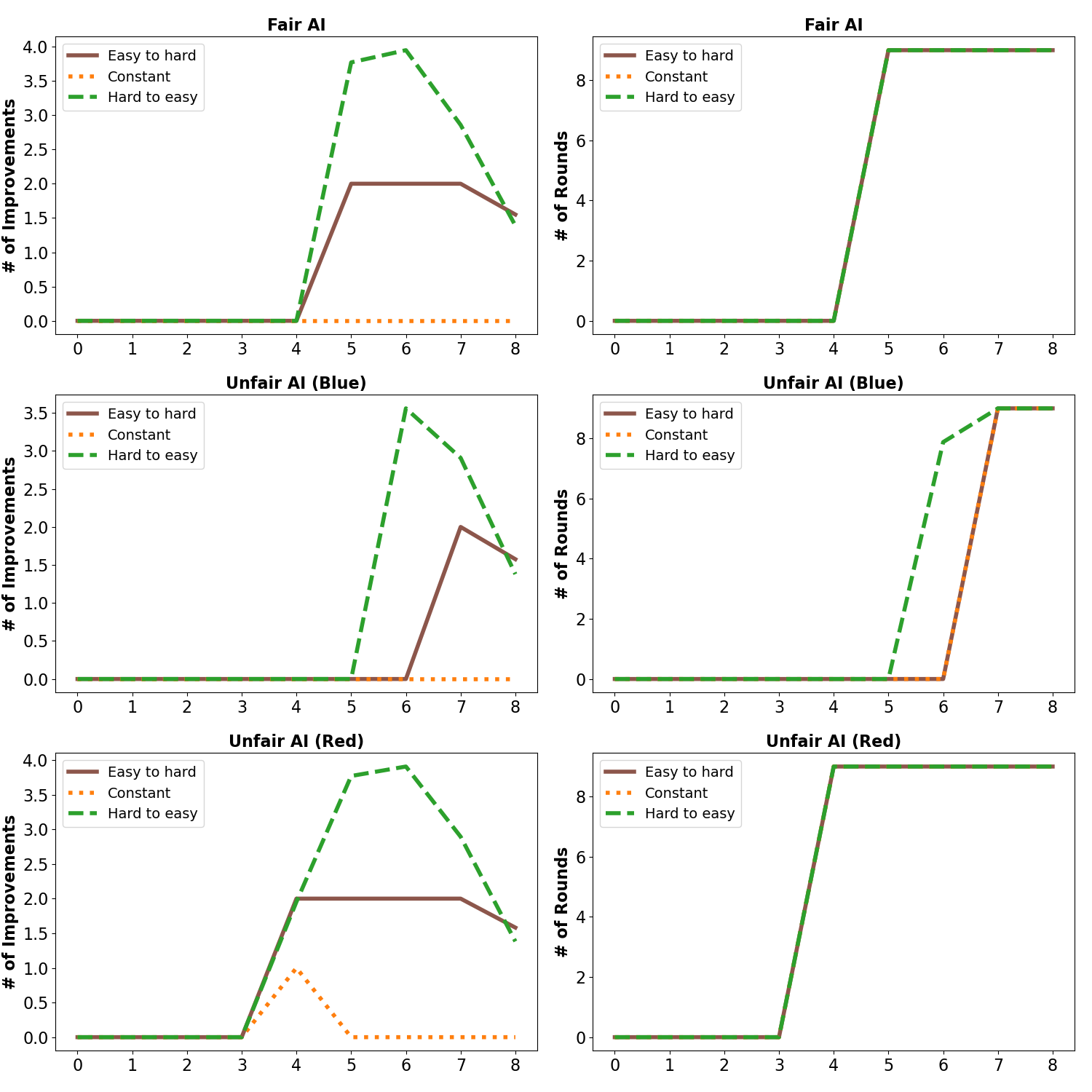}
    \caption{This figure presents the outcomes of implementing a 20-coins improvement cost, considering both Fair and Unfair AI treatments across three studies. The x-axis represents the initial credit score ranges, categorized from 0 to 8, with higher scores indicating higher ranges. In the Fair AI scenario (first row), the figure displays two graphs: the left graph indicates the expected number of improvements, while the right graph shows the expected number of rounds for subjects to keep interacting with the AI model. These are given within each credit score range and for each improvement difficulty used across our three studies. The middle and third rows represent the Unfair AI treatment for blue/female (disadvantaged) and red/male (advantaged) group subjects, similarly displaying ``number of times to improve'' and ``number of rounds to apply for a loan'' strategies as in the first row. The various improvement difficulties are color-coded: constant difficulty used in Studies 1 and 3 (orange), hard to easy in Study 2's sub-experiment (green), and easy to hard in Study 2's other sub-experiment (brown), illustrating different expected strategies for each improvement difficulty across credit score ranges. A high improvement cost might cause most decision subjects to not improve.
    %5 coins improvement cost seems to be a better fit for our setting. The figure highlights the convergence of strategies as the improvement cost increases.
    }
    \label{fig:20coins}
    \vspace*{\fill}
\end{figure*}

Figure~\ref{fig:1coins}--\ref{fig:20coins} show our simulation results for $x=1, 5, 10, 20$, respectively. As we can see, when the cost of qualification improvement is very low (e.g., $x=1$), subjects assigned to the advantaged group in the unfair AI treatment (constant scheme) almost always attempt to improve their qualification in every round (except for subjects who start with the lowest two credit levels, who will directly exit the game). On the other hand, when the cost of qualification improvement is relatively high (e.g., $x=10$ or $x=20$), subjects assigned to the disadvantaged group in the unfair AI treatment (constant scheme) never attempt to improve. However, when we choose an intermediate level of cost (e.g., $x=5$), the number of times subjects would engage with qualification improvement is more responsive to their initial credit level when subjects were placed at an advantaged position by the AI model. Meanwhile, when subjects were placed at a disadvantaged position by the AI model, subjects from more initial credit levels were willing to make improvement attempts. Based on these considerations, we chose 5 coins as the qualification improvement cost in our experiment eventually.

% For 1 coins

\clearpage % Start a new page

% For 5 coins

\clearpage % Start a new page
\newpage
% For 20 coins
\clearpage % Start a new page
\newpage
% For 35 coins

\clearpage
\newpage